\documentclass[12pt]{iopart}
\usepackage{iopams}
\usepackage{amsfonts}
\usepackage{amssymb}
\usepackage{latexsym}
\usepackage{graphicx}

\begin{document}
\title{Recent developments in photon-level operations on travelling light fields}
\author{M. S. Kim }

\address{School of Mathematics and Physics, Queen's University,
Belfast BT7 1NN, United Kingdom}

\date{\today}

\begin{abstract}

Annihilating and creating a photon in a travelling light field are useful building blocks for
quantum-state engineering to generate a photonic state at will.  In
this paper, we review the relevance of these operations to some of the fundamental aspects of quantum physics and recent advances in this
research.
\end{abstract}


\maketitle

\section{Introduction}
We have witnessed a proliferate growth in theoretical and experimental efforts to understand and control physical systems in a quantum level.  In particular, possibilities to apply quantum mechanics toward the radical improvement of information technology have magnetized attention from various branches of physics and beyond. Ever since the advent of lasers, optics has been closely related to the foundations and applications of quantum physics because of its controllability to a very fine level.  However, photons do not interact each other, which makes it difficult to engineer their states.  In order to overcome this problem, various schemes have been suggested.

In a quantum optics laboratory, Gaussian states, whose phase properties are described by Gaussian probability-like functions, were generated but there was some limitation to use them for various tasks of quantum information processing.  There have been suggestions and realisations to engineer the quantum state by subtracting or adding photons from/to a Gaussian field.  Photon subtraction and addition to a given state are plausible ways to manipulate a quantum state.

Many of the quantum optics textbooks start with the definition of
annihilation $\hat{a}$ and creation $\hat{a}^\dag$ operators.  When
they are applied to $|n\rangle$ which represents a state with $n$
number of photons:
\begin{equation}
\hat{a}|n\rangle=\sqrt{n}|n-1\rangle~~~\mbox{and}~~~\hat{a}^\dag|n\rangle=\sqrt{n+1}|n+1\rangle.
\label{1}
\end{equation}
It is easy to recognize that by annihilating (creating) a photon the
photon number changes to $n-1$ ($n+1$) but it is not straightforward
to acknowledge the coefficient $\sqrt{n}$ ($\sqrt{n+1}$).

In 1924, Bose~\cite{bose} found that the most chaotic light bears the
spectrum of Planck's formula at a given temperature under the two
assumptions 1) light is composed of indistinguishable particles and
2) any number of particles can occupy one quantum state. This
is the first full quantum-mechanical explanation of the Planck's
formula for the blackbody radiation. Einstein extended Bose's idea to the quantum theory of an
ideal gas~\cite{einstein}.  As a whole, the particles which bear the
statistics under Bose's assumptions are now called as `bosons'.  Bose's first assumption is against his contemporary belief that particles are always distinguishable.  In the theory of
identical particles, Bose's second assumption naturally brings about
the symmetric nature for the state of bosons in contrast to
anti-symmetric states to fermions.

Let us assume three photons labelled as 1,2 and 3 in the same state
$\psi$.  The three photon state can then be written as
\begin{equation}
|\Psi\rangle=|\psi_1,\psi_2,\psi_3\rangle. \label{symmetric}
\end{equation}
In the Fock state representation, this state is denoted by $|3\rangle$. Subtracting a photon from this three photon state is due to subtracting either photon 1, 2 or 3.  This means that the
subtraction operation is represented by a symmetric operator $\hat{f}=\hat{f}_1+\hat{f}_2+\hat{f}_3$
where subtracting photon $i$ is represented by $\hat{f}_i$. Applying
$\hat{f}$ onto $|\Psi\rangle$, we obtain
\begin{equation}
\hat{f}|\Psi\rangle=|\psi_2,\psi_3\rangle+|\psi_1,\psi_3\rangle+
|\psi_1,\psi_2\rangle
\end{equation}
which is not normalised.  After normalisation
\begin{equation}
\hat{f}|\Psi\rangle=\sqrt{3}\left[{1\over\sqrt{3}}(|\psi_2,\psi_3\rangle+|\psi_1,\psi_3\rangle+
|\psi_1,\psi_2\rangle)\right]
\end{equation}
The state described in the square bracket is the state of two photons, {\it
i.e.} $|2\rangle$ in the Fock representation.  We thus find that by the subtraction of a photon, the state becomes
\begin{equation}
\hat{f}|\Psi\rangle=\sqrt{3}|2\rangle,
\end{equation}
where $\sqrt{3}$ is the coefficient required by the annihilation operation
in (\ref{1}) when $n=3$ for the state in Eq.~(\ref{symmetric}).  It is clear that the coefficient $\sqrt{3}$ appears
naturally from the argument of the symmetric nature of bosonic particles.

The coefficient of $\sqrt{n}$ appears in Eq.(\ref{1}), in fact, tells us that `{\it the probability of a transition in which a boson is absorbed ... is proportional to the number of bosons originally in the state}' as Dirac writes~\cite{dirac}. Similarly we can prove that the creation of a photon in state $|n\rangle$ gets the coefficient $\sqrt{n+1}$ as in Eq.(\ref{1}) purely using the symmetric nature of the bosonic particles.

On the other hand, we know that in harmonic oscillator, whose dynamics has a periodic nature like a wave, the lowering and raising operators, respectively to go down and up the energy ladder also have the same mathematical structure as in Eq.~(\ref{1}) (see for example, Chapter 6 of Ref.~\cite{ballentine}).    Observing that the two are described by the same mathematical formulae but looked at from two different points of view, Dirac writes this equivalence as '{\it one of the most fundamental results of quantum mechanics, which enables a unification of the wave and corpuscular theories of light to be effected.}'~\cite{dirac}.

We can easily derive the commutation relation between the annihilation and creation operators using Eq.(\ref{1}):
$
\hat{a}^\dag\hat{a}\neq \hat{a}\hat{a}^\dag~~~\mbox{but}~~~[\hat{a}, \hat{a}^\dag]=1.
$
In this tutorial, we show how to realise the operation of annihilation and creation of a photon. We then show how to engineer a quantum state using the properties of annihilation and creation operations.

\section{Brief review of quantum optics}
\label{section:2}
We start with a brief review in order to make this tutorial as self-contained as possible.
The annihilation and creation operators are not Hermitian, {\it i.e.} $\hat{a}^\dag\neq\hat{a}$, as is clear in Eq.(\ref{1}), which means that these operators are not measurable.  Thus we define quadrature operators as linear combinations of the two operators
\begin{equation}
\hat{q}=\hat{a}+\hat{a}^\dag~~~,~~~ \hat{p}=i(\hat{a}^\dag-\hat{a})
\label{3}
\end{equation}
which are Hermitian and measurable.  The $m$-th order moment of $\hat{p}$ is
\begin{equation}
 \hat{p}^m = i^m\frac{\partial^m}{\partial v^m}\mbox{e}^{-iv\hat{p}}\mid_{v=0}.
\label{4}
\end{equation}
Similarly, we can find $\hat{q}^m$.  By defining the characteristic function for a field with its density operator $\hat{\rho}$ as
\begin{equation}
C(u,v)=\mbox{Tr}[\hat{\rho}\mbox{e}^{-iv\hat{p}+iu\hat{q}}]
\label{5}
\end{equation}
we can calculate the $m$-th order moments of $\hat{p}$ and $\hat{q}$ at once:
\begin{equation}
\langle\hat{p}^m\rangle=i^m\frac{\partial^m}{\partial v^m}C(u,v)\mid_{u=v=0}~~~,~~~
\langle\hat{q}^m\rangle=(-i)^m\frac{\partial^m}{\partial u^m}C(u,v)\mid_{u=v=0}.
\end{equation}
The characteristic function $C(u,v)$ is called the Weyl characteristic function.

In quantum optics, the exponential operator in Eq.(\ref{5}) is called as the displacement operator, $\hat{D}(\xi)$:
\begin{equation}
\hat{D}(\xi)=\exp(-iv\hat{p}+iu\hat{q})=\exp(\xi\hat{a}^\dag-\xi^*\hat{a})
\label{6}
\end{equation}
where the complex number $\xi=(v+iu)$.  In phase space composed of two conjugate variables $q$ and $p$, $\hat{D}(\xi)$ displaces a state by (Re[$\xi$], Im[$\xi$]) hence the name~\cite{glauber01}:
\begin{equation}
\hat{D}^\dag(\xi)\hat{a}\hat{D}(\xi)=\hat{a}+\xi.
\label{displace-ope}
\end{equation}

In probability theory, the characteristic function is Fourier-transformed to give a probability density function to tell us about the probability of finding a system at the certain value of $p$ and $q$.  However, because of the uncertainty principle in quantum mechanics, it is prohibited to simultaneously allocate certain measurement values to conjugate variables.  Thus, in quantum mechanics, the probability density cannot be defined in phase space.  By the Fourier transformation of the Weyl characteristic function (\ref{5}), we get a function which is like a probability density function so to call it a quasiprobability function~\cite{glauber02}:
\begin{eqnarray}
W(\alpha)&=&{1\over \pi^2}\int C(u,v)\mbox{e}^{-2i\alpha_r u+2i\alpha_i v}dudv \nonumber \\
&=&{1\over \pi^2}\int C(\xi)\mbox{e}^{\alpha\xi^*-\alpha^*\xi}d^2\xi,
\label{Wigner}
\end{eqnarray}
where subscripts $_r$ and $_i$ refer to the real and imaginary parts (this convention is used throughout the paper).  The quasiprobability function was first introduced by Wigner ~\cite{wigner} as an effort to define a probability-like function in phase space.  He derived it under the condition that the marginal probabilities $\int W(p,q)dq$ and $\int W(p,q)dq$ of the probability-like function $W(p,q)$ has to be the true probability functions for the respective variables $p$ and $q$. The Wigner function has a one-to-one correspondence to the density operator and identifies a physical state uniquely.  In practical reasons, the Wigner function is very often used to represent the state of a physical system.

The vacuum state, which is the energy ground state, of a field is denoted by $|0\rangle$.  Using Eq.(\ref{Wigner}), we can find its Wigner function in the Gaussian function.  By applying the displacement operator $\hat{D}(\beta)$ on $|0\rangle$, the peak of the Wigner function is moved by $\beta$ and we obtain
\begin{equation}
W_{|\beta\rangle}(\alpha)= {2\over \pi}\exp[-2|\alpha-\beta|^2]=W_{|0\rangle}(\alpha-\beta),
\label{Wigner-0}
\end{equation}
where $W_{|0\rangle}(\alpha)$ is the Wigner function for the vacuum.
Glauber~\cite{glauber03} introduced the displaced vacuum to analyse the higher-order coherence and called such the state as the coherent state: $|\beta\rangle=\hat{D}(\beta)|0\rangle$.  An ideal laser is assumed to be in a coherent state even though there is some dispute because the phase of a laser field is completely unknown (See \cite{molmer}).

As an example, let us consider to measure the quadrature phase of a superposition of two coherent states with $\pi$ phase difference:
\begin{equation}
|\Psi\rangle=\sqrt{\frac{1}{2(1+\cos\phi_{sch}\mbox{e}^{-2\beta^2})}}(|\beta\rangle+ \mbox{e}^{i\phi_{sch}}|-\beta\rangle),
\label{superposition}
\end{equation}
where $\beta$ is assumed to be real.
This state has been studied in conjunction with the Schr\"{o}dinger paradox~\cite{sch} (See \cite{bu-knight} for a review) and will be discussed further in the latter Sections.  As an example, when $\phi_{sch}=0$, the Wigner function of the coherent superposition state is calculated as
\begin{equation}
W(\alpha)=\frac{1}{\pi(1+\mbox{e}^{-2\beta^2})}\left[\mbox{e}^{-2|\alpha-\beta|^2}+ \mbox{e}^{-2|\alpha+\beta|^2}+2\mbox{e}^{-2|\alpha|^2}\cos(4\beta\alpha_i)\right].
\label{superposition-wigner}
\end{equation}
It is clear that the Wigner function can be negative at some points of the phase space, which confirms that the Wigner function is not a probability function.  In fact, the negativity in phase space is considered to be a signature of nonclassicality for a state.  There have been attempts to observe the negativity in experiment, which will be discussed later.

Another important state of light to mention is a squeezed state.
By a $\chi^{(2)}$ nonlinear interaction, a photon in the input pump field is converted into two photons conserving the energy and momentum.  Arranging the phase-matching condition, the two photons can be emitted into the same mode $a$, in which case the interaction Hamiltonian is $\hat{H}_{s1}=\hbar(\lambda_1^*\hat{a}^2+\lambda_1\hat{a}^{\dag^2})$.  The strength of the pump laser and the nonlinear coefficient determine the coupling parameter $\lambda_1$.  On the other hand, if the two daughter photons are emitted into two distinctive modes $a$ and $b$, the interaction Hamiltonian becomes $\hat{H}_{s2}=\hbar(\lambda_2^*\hat{a}\hat{b}+\lambda_2\hat{a}^{\dag}\hat{b}^\dag)$.  The dynamics of the field in modes $a$ and $b$ can be obtained by applying the evolution operator depending on the Hamiltonian.  Assuming that the field is initially prepared with nothing, {\it i.e.} in the vacuum state $|0\rangle$, at the interaction time $t$ it transforms into
\begin{equation}
|{\cal S}_1\rangle\equiv\hat{S}_1(\zeta)|0\rangle\equiv\mbox{e}^{-i\hat{H}_{s1}t/\hbar}|0\rangle~~~ ;~~~\zeta=2i\lambda_1 t
\label{sq1}
\end{equation}
in the case of single-mode emission. The evolution operator $\hat{S}_1(\zeta)$ is called the single-mode squeezing operator and the state in Eq.(\ref{sq1}) is called the squeezed state (or, more precisely, squeezed vacuum).  There may be other types of squeezed states by applying the squeezing operator to various initial states.  In this paper, however, we will be interested mainly in the squeezed vacuum.  After the decomposition of the squeezing operator, we can calculate the precise form of the squeezed vacuum in the Fock state basis (See the derivation in \cite{caves1} and Appendix 5 in \cite{barnett-radmore}):
\begin{equation}
|{\cal S}_1\rangle=\sqrt{\mbox{sech} \zeta}\sum_{n=0}^\infty\frac{\sqrt{(2n)!}}{n!}(-{1\over 2}\tanh \zeta)^n|2n\rangle,
\label{sq1-state}
\end{equation}
where $\zeta$ is considered to be real for simplicity.  It is
immediately recognised that there are only even numbers of photons to
be realised in this state, which reflects the nature of twin photon
generation by the $\chi^{(2)}$ nonlinear interaction.  The  unitary
transformation of the bosonic operator is
\begin{equation}
\hat{S}_1(\zeta)\hat{a}\hat{S}_1^\dag(\zeta)=\hat{a}\cosh \zeta+\hat{a}^\dag\sinh \zeta.
\label{unitary-s1}
\end{equation}

If two photons are emitted into two distinctive modes, the vacuum state evolves into
\begin{equation}
|{\cal S}_2\rangle\equiv \hat{S}_2(\zeta)|0,0\rangle \equiv \mbox{e}^{-i\hat{H}_{s2}t/\hbar}|0,0\rangle
= \mbox{sech} \zeta\sum_{n=0}^\infty (-\tanh \zeta)^n|n,n\rangle,
\label{sq2}
\end{equation}
where the evolution operator is now the two-mode squeezing operator $\hat{S}_2(\zeta)$ with the squeezing parameter $\zeta=i\lambda_2t$.
Again, the decomposition of the two-mode squeezing operator $\hat{S}_2(\zeta)$ has been used and $\zeta$ has been assumed real for simplicity.  We observe in the two-mode squeezed state (\ref{sq2}) that if there are $n$ photons in mode $a$, we can say that there are also $n$ photons in mode $b$ without having to measure it.  This `deterministic nature' of the quantum state is an important ingredient of quantum correlation, so-called {\it entanglement}, the detail of which will be further discussed later.
The unitary transformations of bosonic operators are
\begin{equation}
\hat{S}_2(\zeta)\hat{a}\hat{S}_2^\dag(\zeta)=\hat{a}\cosh \zeta+\hat{b}^\dag\sinh \zeta~~~;
~~~\hat{S}_2(\zeta)\hat{b}\hat{S}_2^\dag(\zeta)=\hat{b}\cosh \zeta+\hat{a}^\dag\sinh \zeta.
\label{unitary-s2}
\end{equation}
The generation of an entangled state from a single-mode squeezed vacuum has been discussed in \cite{kim01} (See a review article \cite{braunstein01}).

Using Eq.~(\ref{sq1-state}) in the definition of the Weyl characteristic function (\ref{5}), it is clear that the squeezing operator transforms the coordinates of phase space, expanding one axis at the expense of contracting the other.  The Wigner function for the single-mode squeezed state is calculated to be
\begin{equation}
W_{S|0\rangle}(\alpha)=W_{|0\rangle}(\alpha_r\mbox{e}^{\zeta}, \alpha_i\mbox{e}^{-\zeta}).
\label{sq1-Wigner}
\end{equation}
The Wigner function thus appears squeezed in phase space hence the name squeezed state.  In fact, the unitary transformations (\ref{unitary-s1}) and (\ref{unitary-s2}) are the linear transformations of the set of initial bosonic operators to that of the output bosonic operators: in other words, we can write the transformation in the following general form with the actual form of the transformation matrix $\mathbf{T}$ to be determined accordingly,
\begin{equation}
\hat{S}_{1,2}(\zeta)\left(\begin{array}{c}
\hat{a} \cr
\hat{a}^\dag \cr
\hat{b} \cr
\hat{b}^\dag \cr
\end{array}\right)\hat{S}_{1,2}^\dag(\zeta)=\mathbf{T}\left(\begin{array}{c}
\hat{a} \cr
\hat{a}^\dag \cr
\hat{b} \cr
\hat{b}^\dag \cr
\end{array}\right)
\label{linear-trans}
\end{equation}
Huang and Agarwal have shown that the initial Gaussian Wigner function remains Gaussian after a linear transformation of bosonic operators of fields~\cite{huang}.  They then showed that the action of a beam splitter, squeezer, linear amplifier are all described by a linear transformation.  With the reasons of mathematical handedness and experimental relevance, a Gaussian state in single-mode and multimode fields have been extensively studied.

Throughout the paper, a homodyne detector is used to measure the properties of the light field.  The quantum theory of homodyne detection was originated by Yuen and Shapiro~\cite{yuen}.  In order to measure quadrature variables, the homodyne detector makes use of interference between a reference field, called a local oscillator, and the field to probe, by superposing them at a beam splitter.  Let us consider a lossless beam splitter of two input ports of modes $a$ and $b$ and two output ports of $a_{out}$ and $b_{out}$. Using the SU(2) symmetry, Campos, Saleh and Teich~\cite{campos} have developed the beam splitter operator
\begin{equation}
\hat{B}(\theta, \phi)=\exp\left\{\frac{\theta}{2}(\hat{a}^\dag\hat{b}\mbox{e}^{i\phi}- \hat{a}\hat{b}^\dag\mbox{e}^{-i\phi})\right\}
\label{beamsplitter}
\end{equation}
with the reflectivity $r=\sin{\theta\over 2}$ and the transmittivity $t=\cos{\theta\over 2}$.  It is important to note that the beam splitter operator is unitary to conserve energy between input and output fields.  Due to the unitary nature, in fact, two identical photons respectively put into two input ports interfere at the beam splitter to exit together at an output port.  This is the basic idea of the so-called Hong-Ou-Mandel interferometer which appears often in quantum technology to prove the identity of two photons~\cite{hong}.  In this paper, we fix the phase difference given by the beam splitter to $\phi=0$, for convenience.  The unitary transformation by a beam splitter is then
\begin{equation}
\left(\begin{array}{c}
\hat{a}_{out} \cr
\hat{b}_{out} \cr
\end{array}\right)
=\left(\begin{array}{cc}
t & -r \cr
r & t \cr
\end{array}\right)
\left(\begin{array}{c}
\hat{a} \cr
\hat{b} \cr
\end{array}\right)
\end{equation}
which is again a linear transformation.

Photon numbers are measured at the beam splitter output ports and the difference between them is
\begin{equation}
\hat{n}_{b}-\hat{n}_a= (t^2-r^2)(\hat{a}^\dag\hat{a}- \hat{b}^\dag\hat{b})+2tr(\hat{a}^\dag\hat{b}+ \hat{a}\hat{b}^\dag),
\end{equation}
where the photon number operators $n_{a}=\hat{a}_{out}^\dag\hat{a}_{out}$ and $\hat{n}_b=\hat{b}_{out}^\dag\hat{b}_{out}$.  For a 50:50 beam splitter with $r=t=1/\sqrt{2}$, the photon number difference is
\begin{equation}
\hat{n}_{a}-\hat{n}_b= \hat{a}^\dag\hat{b}+ \hat{a}\hat{b}^\dag.
\end{equation}
The homodyne detector with the 50:50 beam splitter is specially called as the balanced homodyne detector.   Now, we take a strong laser field as the local oscillator which means the local oscillator can be considered as a classical field to replace the bosonic operators with its dimensionless amplitude $\mathcal{E}_L$ and phase $\phi_L$: $\hat{b}\sim|\mathcal{E}|\mbox{e}^{i\phi_L}$ and $\hat{b}^\dag\sim|\mathcal{E}_L|\mbox{e}^{-i\phi_L}$. This approximation brings
\begin{equation}
\hat{n}_{b}-\hat{n}_a= |\mathcal{E}_L|(\hat{a}^\dag\mbox{e}^{i\phi_L}+ \hat{a}\mbox{e}^{-i\phi_L})\equiv\hat{N}_{\phi_L}.
\label{homodyne-theta}
\end{equation}
It is easily seen that the photon number difference is the scaled quadrature operator.  When $\phi_L=0$,
$\hat{N}_{0}=|\mathcal{E}_L|\hat{q}$.  When $\phi_L=\pi/2$, $\hat{N}_{\pi/2}=|\mathcal{E}_L|\hat{p}$.  Now, it is clear that quadrature operators can be measured by the homodyne detector.  In fact, by changing $\phi_L$, we can measure not only two conjugate operators $\hat{p}$ and $\hat{q}$ but also the  phase value along any rotated axis in phase space.  By defining the quadrature operator $\hat{q}_\phi=(\hat{a}\mbox{e}^{-i\phi}+\hat{a}^\dag\mbox{e}^{i\phi})$, we have the correspondence of the measured value to the quadrature variable: $\hat{N}_{\phi_L}=|\mathcal{E}_L|\hat{q}_\phi$.

According to the probability theory, the probability of photon number difference is the Fourier transformation:
\begin{equation}
P(N_{\phi_L})={1\over 2\pi}\int dk C_{\phi_L}(k)\mbox{e}^{-ikN_{\phi_L}}
\label{p}
\end{equation}
where
\begin{equation}
C_{\phi_L}=\mbox{Tr}\left[\hat{\rho}\mbox{e}^{ik|\mathcal{E}_L|(\hat{a}\mbox{e}^{-i\phi_L}
+ \hat{a}^\dag\mbox{e}^{i\phi_L})}\right].
\end{equation}
Comparing Eq.(\ref{p}) with Eq.(\ref{Wigner}), we can find that $P(N_{\phi_L})$ is proportional to the marginal Wigner function.  The probability of photon number difference is the marginal Wigner function whose axis is determined by the phase of the local oscillator.  With this observation, Risken and Vogel proposed to reconstruct the Wigner function using the homodyne measurement outcomes by rotating the local oscillator phase~\cite{vogel}.  This is the so-called optical tomography.

The homodyne detector is also known for its high efficiency.
Without approximating the local oscillator as a classical field, Braunstein assessed the reliability of the homodyne measurement as a measurement of quadrature phase, depending on the intensity of the local oscillator~\cite{braunstein-hom}.  He found that the strength of the local oscillator is not sufficient to ensure that a homodyne detector acts like an ideal detector of quadrature phase.

Before we finish this section, let us go back to the Wigner function. As it is not possible to define a probability density function in phase space, we had to define a quasiprobability function which is a probability like function.  As it is not the true probability function, it is not surprising to have more than one quasiprobability function~\cite{agarwal-quasi}.  Using the Campbell-Baker-Hausdorf theorem\footnote{If $[\hat{A},[\hat{A},\hat{B}]]=[\hat{B},[\hat{A},\hat{B}]]=0$, then $\exp(\theta_1\hat{A}+\theta_2\hat{B})=\exp(\theta_1\hat{A})\exp(\theta_2\hat{B}) \exp(-[\theta_1\hat{A},\theta_2\hat{B}]/2)$.-See, for example, p.42-44 of \cite{barnett-radmore} for a proof.} the displacement operator in Eq.~(\ref{6}) can be written as
\begin{equation}
D(\xi)=\mbox{e}^{\xi\hat{a}^\dag}\mbox{e}^{-\xi^*\hat{a}}\mbox{e}^{-|\xi|^2/2} =\mbox{e}^{-\xi^*\hat{a}}\mbox{e}^{\xi\hat{a}^\dag}\mbox{e}^{|\xi|^2/2}
\label{displacem-cha}
\end{equation}
The Weyl characteristic function Eq.~(\ref{5}) can then be written as
\begin{equation}
C(\xi)=\mbox{Tr}\left[\mbox{e}^{\xi\hat{a}^\dag}\mbox{e}^{-\xi^*\hat{a}}\hat{\rho}\right]\mbox{e}^{-|\xi|^2/2}
=\mbox{Tr}\left[\mbox{e}^{-\xi^*\hat{a}}\mbox{e}^{\xi\hat{a}^\dag}\hat{\rho}\right]\mbox{e}^{|\xi|^2/2}
\label{character-2}
\end{equation}
from which we can see that the expectation value of the normally ordered operators
\begin{equation}
\langle\hat{a}^{\dag^{n}}\hat{a}^m\rangle=\left(\frac{\partial}{\partial \xi}\right)^m \left(-\frac{\partial}{\partial \xi^*}\right)^n C(\xi)\mbox{e}^{\frac{1}{2}|\xi|^2}
\label{normal-order}
\end{equation}
and that of the antinormally ordered operators is
\begin{equation}
\langle\hat{a}^{m}\hat{a}^{\dag^{n}}\rangle=\left(\frac{\partial}{\partial \xi}\right)^m \left(-\frac{\partial}{\partial \xi^*}\right)^n C(\xi)\mbox{e}^{-\frac{1}{2}|\xi|^2}.
\label{anti-normal-order}
\end{equation}
This shows that by Fourier transformation of $C(\xi)\mbox{e}^{\pm|\xi|^2/2}$, we will have probability-like functions related to the expectation values of normally and antinormally ordered operators.  In general, we define $s$-parameterised characteristic function and quasiprobability functions as
\begin{equation}
C(\xi,s)=C(\xi)\mbox{e}^{{s\over
2}|\xi|^2}=\mbox{Tr}[\hat{D}(\xi,s)\hat{\rho}]~~;~~
\hat{D}(\xi,s)\equiv\hat{D}(\xi)\mbox{e}^{{s\over 2}|\xi|^2}
\label{character-s}
\end{equation}
where $-1\leq s\leq 1$ and
\begin{equation}
P(\alpha,s)={1\over\pi^2}\int_{-\infty}^{\infty}d^2\xi C(\xi,p)\exp(\alpha\xi^*-\alpha^*\xi).
\label{quasi-s}
\end{equation}
It is straightforward to show that $P(\alpha,s)$ satisfies one of the conditions for it to be a probability function: $\int_{-\infty}^{\infty} P(\alpha,s)d^2\alpha=1$ for any $s$.  However, $P(\alpha,s)$ does not have to be positive according to the definition.

When $s=0$, $P(\alpha,s)$ becomes the Wigner function.  When $s=1$,
$P(\alpha,s)$ refers to the quasiprobability for normally ordered
operators and called as the Glauber-Sudarshan $P$ function to honour
its inventors~\cite{glauber01,sudarshan} (Throughout the paper, the
Glauber Sudarshan $P$ function is referred by the $P$ function for
simplicity). It has been shown that the $P$ function is related to
the density operator through the following equation
\begin{equation}
\hat{\rho}=\int_{-\infty}^{\infty}P(\alpha)|\alpha\rangle\langle\alpha|d^2\alpha.
\label{p-function}
\end{equation}
It is known that any moments of photon number operators should be normally ordered~\cite{loudon}, for example
\begin{equation}
\bar{n}=\langle\hat{n}\rangle~~~; ~~~\overline{n^2}=\langle:\hat{n}^2:\rangle = \langle\hat{a}^\dag\hat{a}^\dag \hat{a}\hat{a}\rangle
\label{photon-number}
\end{equation}
where $:~:$ stands for normal-ordering.  The photon number variance is then
\begin{equation}
\langle :\Delta\hat{n}^2:\rangle=\langle :\hat{n}^2:\rangle-\langle\hat{n}\rangle^2
\label{photon-variance}
\end{equation}
If the $P$ function is positive then $P(\alpha)$ can be considered
as a probability function and the normally ordered variance will be
positive.  However, if $P$ function is not a well-behaved positive
function, the normally ordered variance does not have to be
positive.  The non-existence of the positive well-behaved $P$
function is thus considered to be a sign of nonclassicality for a
given field.

Now, we have seen two criteria of nonclassicality: one
is based on negativity of the Wigner function and the other is based
on the non-existence of the well-behaved $P$ function.  These two
conditions do not necessarily coincide so there seems to be a subtle
problem to be settled.  In fact, as we can show later, the
positivity of $P$ function guarantees the positivity of the Wigner
function but the converse is not necessarily true.

We can obtain $P(\alpha,s^\prime)$ from another quasiprobability function $P(\alpha,s)$ for $s^\prime<s$~\cite{barnett-n}:
\begin{equation}
P(\alpha,s^\prime)=\frac{2}{\pi(s-s^\prime)}\int_{-\infty}^{\infty}d^2\beta \exp\left( \frac{2|\alpha-\beta|^2}{s-s^\prime} \right)P(\beta,s),
\label{s-s}
\end{equation}
which appears as a convolution of $P(\alpha,s)$ with a Gaussian function and its effect is to bring the peaks of $P(\alpha,s^\prime)$ broader than those of $P(\alpha,s)$ (In fact, this broadening reminds us of a decoherence process of the field and the detail has been studied using the Gaussian convolution theory~\cite{kim-imoto}).  The $s$-parameterised quasiprobability function (\ref{quasi-s}) together with (\ref{character-s}) can be written as
\begin{equation}
P(\alpha,s)=\mbox{Tr}[\hat{D}^\dag(\alpha)\hat{\rho}\hat{D}(\alpha)\hat{T}(s)]
\label{p-t}
\end{equation}
where $\hat{T}(s)={1\over\pi^2}\int d^2\xi \hat{D}(\xi,s)$. It can be shown for $s<1$ that~\cite{barnett-radmore}
\begin{equation}
\hat{T}(s)={1\over\pi(1-s)}:\exp\left(-{2\over 1-s}\hat{a}^\dag\hat{a}\right):
\label{T-aa}
\end{equation}
Then expanding the exponential function and using the completeness $\sum_{k=0}^{\infty}|k\rangle\langle k|=\mathbf{1}$, we find that
\begin{eqnarray}
&&:\exp\left(-{2\over 1-s}\hat{a}^\dag\hat{a}\right): \nonumber \\
&&= 1-{2\over 1-s}\hat{a}^\dag\hat{a} +{1\over 2!}\left({2\over 1-s}\right)^2\hat{a}^{\dag^2}\hat{a}^2- {1\over 3!}\left({2\over 1-s}\right)^3\hat{a}^{\dag^3}\hat{a}^3+\cdots \nonumber \\
&&=\sum_{k=0}^\infty\sum_{n=0}^k\left(-{2\over 1-s}\right)^n\left(\begin{array}{c}
k \cr
n \cr
\end{array}\right)|k\rangle\langle k| =\sum_{k=0}^\infty \left({s+1\over s-1}\right)|k\rangle\langle k|.
\end{eqnarray}
Substituting this into Eq.~(\ref{p-t}), the $s$-parameterised quasiprobability function can be written as
\begin{equation}
P(\alpha,s)={2\over\pi(1-s)}\sum_{k=0}^\infty \left(\frac{s+1}{s-1} \right)^k\langle k|\hat{D}^\dag(\alpha) \hat{\rho}\hat{D}(\alpha)|k\rangle.
\label{quasi-density}
\end{equation}
which allows us to have a connection between the quasiprobability functions and the density matrix elements.  In fact, through this connection, we can also reconstruct a quasiprobability function after the density matrix is reconstructed from homodyne measurements (See \cite{ulf2} for details).

When $s=0$, $P(\alpha,s)$ in Eq.~(\ref{quasi-density}) becomes the Wigner function:
\begin{equation}
W(\alpha)={2\over\pi}\sum_{k=0}^\infty (-1)^k\langle k|\hat{D}^\dag(\alpha) \hat{\rho}\hat{D}(\alpha)|k\rangle
\label{Wigner-2}
\end{equation}
The Wigner function at $\alpha$ of phase space is the probability of
having an even number of photons less that of having an odd number
of photons after displacing the field by $\alpha$.  The dependence
of the Wigner function on the even and odd parities is important for
its reconstruction and applications.  When $s=-1$, $P(\alpha,s)$ is
called the $Q$ function, which is the quasiprobability function
related to the expectation values of antinormally ordered operators.
From Eq.~(\ref{quasi-density}), we can find that $Q(\alpha)$ is
the overlap between the field and the coherent state
$|\alpha\rangle$: \begin{equation}
Q(\alpha)={1\over\pi}\langle\alpha|\hat{\rho}|\alpha\rangle.
\label{q-function}
\end{equation}
Thus the $Q$ function is positive at any point of phase space.  At one end $s=-1$, the $s$-parameterised quasiprobability function is positive well-behaved for any state.  As $s$ grows, the peaks of the function become sharper.  At the other end $s=1$, the $s$-parameterised quasiprobability function may not even exist depending on the state of the field.

For quantum state reconstruction, the homodyne measurement with dark count and other sources of inefficiency have to be assessed.  The optimum number of measurement angles also has to be assessed. After the data is collected, the quantum state reconstruction has to be considered with the algorithm of  maximum likelihood~\cite{ulf2}.

\section{Photon subtraction and addition using a cavity or a beam splitter}
\label{add-subtract}
While photon addition and subtraction have been discussed for long, neither of the operations were experimentally demonstrated till recently.  Wenger, Tualle-Brouri and Grangier~\cite{grangier01} were able to subtract a single photon from a travelling field using a beam splitter and a photodetector.  In the same year, a single-photon addition was experimentally performed by Zavatta, Viciani and Bellini~\cite{bellini01}.  The combination of the subtraction and addition has been successfully demonstrated in \cite{bellini-kim}.

\begin{figure}
\centering{\includegraphics{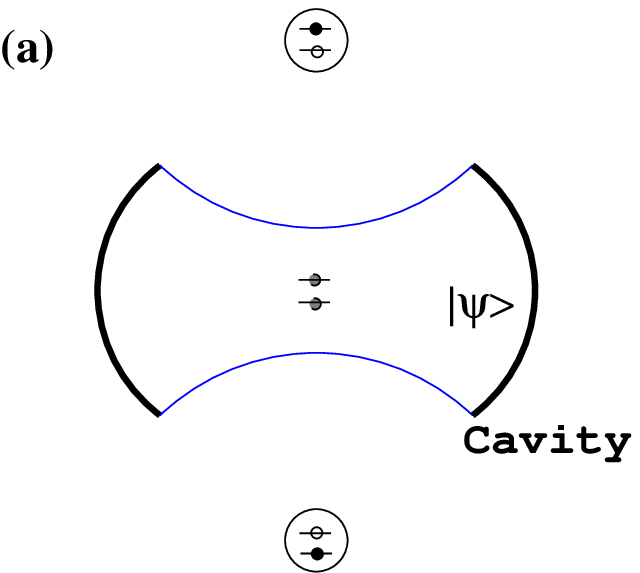}}
\hspace*{0.5cm}
\centering{\includegraphics{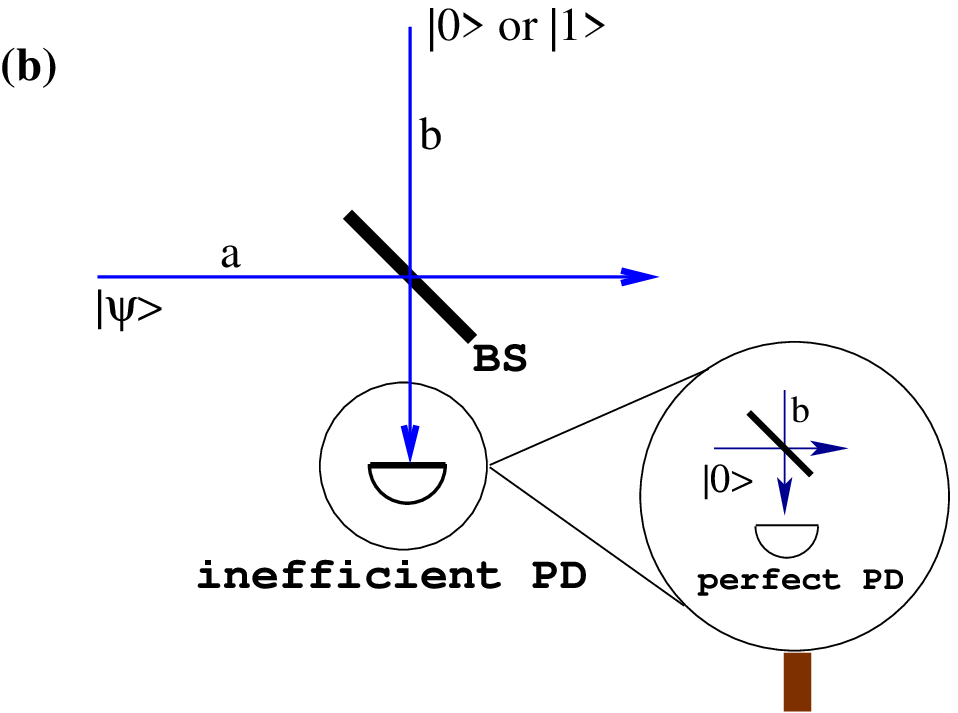}}
\caption{{\bf(a)}:Operations on a field by passing a two-level atom through a cavity.  The atom is initially prepared in its ground state and then measured
in its excited state at the exit in order to subtract a photon (filled dots in the level scheme).
The atom is initially prepared in its excited state and then measured in its ground state at
the exit in order to add a photon (empty dots in the level scheme); {\bf (b)}: Operations on a field by using
a high-transmittivity beam splitter.  A photon is subtracted conditioned on the vacuum input to take
away a single photon and detected as a single photon state at the output.
If a single-photon state is input to the beam splitter and leaves it as a vacuum,
a single photon has been added to the field input to the other input port of the beam splitter. In the
inset, the equivalence of an imperfect photodetection with its efficiency $\eta$ is depicted using
a perfect photodetector with a beam splitter of transmittivity $\eta$ in front. BS: beam splitter. PD: photodetector}
\label{fig:add-sub}
\end{figure}

In order to illustrate how to add and subtract a photon, let us consider a perfect single-mode cavity where a state of a field has been prepared (see Fig.~\ref{fig:add-sub} (a)).  We then send a two-level atom of its excited $|e\rangle$ and ground $|g\rangle$ states through the cavity.  Assuming resonant interaction between the atom and the cavity field, the Hamiltonian in the interaction picture is written as
\begin{equation}
\hat{H}_{jc}=\hbar\lambda_{jc}(\hat{a}^\dag\hat{\sigma}_-+\hat{a}\hat{\sigma}_+),
\label{int-hamiltonian}
\end{equation}
where $\lambda_{jc}$ is the coupling constant, $\hat{\sigma}_+=|e\rangle\langle g|$ and $\hat{\sigma}_-=|g\rangle\langle e|$.  This fully-solvable Hamiltonian has been widely studied under the name of Jaynes-Cummings model (See \cite{knight-jcm} for a review).  Let us assume that the atom is injected into the cavity in its excited state then it is measured in its ground state after interaction with the cavity field at the exit of the cavity.  This is described by
\begin{equation}
\hat{A}_{jc}\equiv\langle g|\mbox{e}^{-i\hat{H}_{jc}t/\hbar}|e\rangle=-i\frac{\sin(\lambda_{jc}\sqrt{\hat{n}}t)}{\sqrt{\hat{n}}}\hat{a}^\dag.
 \label{addition-cavity}
\end{equation}
where the orthonormality $\langle g|e\rangle=0$ has been used.
When $\frac{\sin(\lambda_{jc}\sqrt{\hat{n}}t)}{\sqrt{\hat{n}}}\approx\mathbf{1}$, the atom will add a photon to the cavity field without any extra operation.  The condition is satisfied when $\lambda_{jc}$ is small and the photon number inside the cavity is small.   Similarly, if the atom is prepared in its ground state and measured in its excited state after the cavity field interaction, a photon is subtracted from the cavity under the same condition.  The operation is described by $\hat{S}_{jc}=-i\hat{a}\frac{\sin(\lambda_{jc}\sqrt{\hat{n}}t)}{\sqrt{\hat{n}}}$.
 In fact, a single photon state has been generated experimentally by sending an atom to an empty cavity (prepared in the vacuum $|0\rangle$)~\cite{haroche-single}.  Note that any constant value in front of $\hat{a}$ or $\hat{a}^\dag$ is neglected because that has to disappear when the final normalisation is taken place.  Throughout the paper, we will ignore those constant values.

However, according to the input-output theory, the photon subtraction and addition assisted by a cavity does not seem to be an optimum solution even though a cavity may be a candidate for other types of light field engineering.   Let us thus take a travelling field in mode $a$ to a beam splitter as shown in Fig.~\ref{fig:add-sub} (b), where nothing has been put into the other input port of mode $b$ ('nothing' means the vacuum state).  If the injected vacuum takes away a single photon, the action of a photon subtraction has been performed.   In order to analyse this action, let us consider the generators $\hat{J}_+=\hat{a}\hat{b}^\dag$ and $\hat{J}_-=\hat{a}^\dag\hat{b}$ of the beam splitter operator $\hat{B}(\theta,\phi=0)$ in Eq.~(\ref{beamsplitter}).  Together with $\hat{J}_3={1\over 2}(\hat{b}^\dag\hat{b}-\hat{a}^\dag\hat{a})$, $\hat{J}_+$ and $\hat{J}_-$ form a group thus we can decompose\footnote{The Schwinger's relation~\cite{schwinger} is to cast the two-dimensional harmonic oscillator in terms of an angular momentum system normalised to $\hbar$. The relation is based on the fact that $\hat{J}_{1}={1\over 2}(\hat{J}_++\hat{J}_-)$, $\hat{J}_2={1 \over 2i}(\hat{J}_+-\hat{J}_-)$ and $\hat{J}_3$ have the same commutation relations as the angular momentum operators.  As the operators form a complete group, it is clear that the decomposition is possible even though the actual decomposition is not necessarily straightforward~\cite{kim-mc}.} the beam splitter operator as (For example, see pages 45-46 in \cite{barnett-radmore})
\begin{equation}
\hat{B}(\theta,0)=\mbox{e}^{-\tan{\theta\over 2}\hat{J}_+}~\mbox{e}^{-\ln(\cos^2{\theta\over 2})\hat{J}_3} ~\mbox{e}^{\tan{\theta\over 2}\hat{J}_-}.
\label{beam-splitter-decom}
\end{equation}
As input mode $b$ is in the vacuum,
\begin{equation}
\hat{B}(\theta,0)|0\rangle_b=\mbox{e}^{-\tan{\theta\over 2}\hat{J}_+}t^{\hat{a}^\dag\hat{a}}|0\rangle_b
\end{equation}
where $t$ is the transmittivity of the beam splitter as in Eq.~(\ref{beamsplitter}).  This is further reduced if we project it onto a single-photon state. Thus the action of single-photon annihilation is summarised as
\begin{equation}
\hat{S}_{bs}=~_b\langle 1|\hat{B}(\theta,0)|0\rangle_b=-\tan{\theta\over 2}\hat{a}t^{\hat{a}^\dag\hat{a}}\sim \hat{a}t^{\hat{a}^\dag\hat{a}}.
\label{sub-bs}
\end{equation}
It is clear that the projection gives an impact of subtracting a photon from the field in mode $a$ but in order to leave only the annihilation operator, the following condition has to be satisfied
\begin{equation}
t^{\hat{a}^\dag\hat{a}}\approx \mathbf{1},
\label{condition-bs}
\end{equation}
which can be compared with the condition leading to the annihilation operation using the Jaynes-Cummings model.
The factor $-\tan{\theta\over 2}$ in Eq.~(\ref{sub-bs}) disappears during normalisation.    According to the definition
of the beam splitter operator, the parameter $\theta$ determines reflectivity and transmittivity, and
$\theta$ small means the reflectivity small.

If, instead of the vacuum, a single-photon state is injected to mode $b$ and nothing is measured at the output, we know that one photon has been added and the overall operation is
\begin{equation}
\hat{A}_{bs}=~_b\langle 0|\hat{B}(\theta,0)|1\rangle_b=\tan{\theta\over 2}t^{\hat{a}^\dag\hat{a}}\hat{a}^\dag\sim t^{\hat{a}^\dag\hat{a}}\hat{a}^\dag.
\label{bs-add-ex}
\end{equation}
Under the condition (\ref{condition-bs}), the projection $\hat{A}_{bs}$ becomes to simulate $\hat{a}^\dag$. However, this scheme is demanding because, the single photon generation on demand is challenging and the detection of the vacuum accompanies higher noise than the detection of photons.

In the photon subtraction scheme using the beam splitter, the most important experimental
challenge is the single-photon detection.  There are many issues in the inefficient detection including
1) the low detection efficiency of a photodetector and 2) the detection of a photon in a wrong frequency band.  An inefficient photodetector with its detection efficiency $\eta$ can be simulated by a combination of a perfect photodetector with a beam splitter in front (See Fig.~\ref{fig:add-sub} (b)), where the vacuum (mode $v$) is assumed to be injected to the unused input and the transmittivity of the beam splitter is the same as the detection efficiency $\eta$.  Assuming the input field to subtract a photon is $\hat{\rho}_a$, the output field in mode $a$ is
\begin{equation}
\hat{\rho}_{s}=\mbox{Tr}_v(~_b\langle 1|\hat{B}_{bv}\hat{B}_{ab}\hat{\rho}_a \otimes\left(|0,0\rangle_{bv} \langle0,0| \right) \hat{B}^\dag_{bv}\hat{B}^\dag_{ab}|1\rangle_b)
\label{sub-bs-ineff}
\end{equation}
Using the expansion of the beam splitter operator (\ref{beamsplitter}) under the condition (\ref{condition-bs}),
\begin{equation}
\hat{\rho}_{s}\approx \left({\theta\over 2}\right)^2\eta^2 \hat{a}\hat{\rho}_a\hat{a}^\dag  +  \left({\theta\over 2}\right)^4\eta^2 (1-\eta^2)\hat{a}^2\hat{\rho}_a\hat{a}^{\dag^2}.
\label{sub-bs-ineff-s}
\end{equation}
The first term represents the annihilation of a single photon while the second term is to annihilate two photons.  For the second term to be negligible, the reflectivity ($\theta$) of the beam splitter has to be small and the detection efficiency has to be large.

Currently, the efficiency of single-photon level detection is
very low, even though it depends on the frequency of the field
to be probed.  On the other hand, an avalanche photodiode can be used to detect photons as it is highly sensitive to a photon.  It is, however, saturated by a single photon, which means that a photodiode is
like a on-off detector to be `on' if there are photons and `off' if there is no photon.  It
cannot tell how many photons there are.  Mathematically, `on' is to project the field onto
\begin{equation}
\hat{G}=\sum_{n=1}^\infty |n\rangle\langle n|=\mathbf{1}-|0\rangle\langle 0|.
\label{on-off}
\end{equation}
Instead of a single-photon detector, the on-off detector is used in recent photon subtraction
experiments~\cite{grangier01,polzik01,grangier02,bellini01}.  In order to see how it works in the annihilation of a photon, let us replace the single-photon detector with $\hat{G}$ in
Eq.~(\ref{sub-bs}).  For the density operator for the input field in mode $a$ is $\hat{\rho}_a$, the output conditioned on an `on' event is
\begin{eqnarray}
\mbox{Tr}_b[\hat{G}_b\hat{B}(\theta,0)\hat{\rho}_a\otimes\left(|0\rangle_b\langle0|\right)\hat{B}^\dag(\theta,0)]
&=&\mbox{Tr}_b[\hat{B}(\theta,0)\hat{\rho}_a\otimes\left(|0\rangle_b\langle0|\right)\hat{B}^\dag(\theta,0))]
\nonumber \\
&-& ~_b\langle 0| \hat{B}(\theta,0)\hat{\rho}_a\otimes\left(|0\rangle_b\langle0|\right)\hat{B}^\dag(\theta,0)|0\rangle_b.
\label{sub-on-off-1}
\end{eqnarray}
where
\begin{equation}
\mbox{Tr}_b[\hat{B}(\theta,0)\hat{\rho}_a\otimes\left(|0\rangle_b\langle0|\right)\hat{B}^\dag(\theta,0))]\approx
t^{\hat{a}^\dag\hat{a}}\hat{\rho}_a t^{\hat{a}^\dag\hat{a}}+\tan^2{\theta\over 2}\hat{a} ~t^{\hat{a}^\dag\hat{a}}\hat{\rho}_a t^{\hat{a}^\dag\hat{a}}~\hat{a}^\dag
\end{equation}
under the condition (\ref{condition-bs}) and
\begin{equation}
~_b\langle 0| \hat{B}(\theta,0)\hat{\rho}_a\otimes\left(|0\rangle_b\langle0|\right)\hat{B}^\dag(\theta,0)|0\rangle_b =
t^{\hat{a}^\dag\hat{a}}\hat{\rho}_a t^{\hat{a}^\dag\hat{a}}
\end{equation}
Substituting the above calculations into Eq.~(\ref{sub-on-off-1}), we find that
\begin{equation}
\mbox{Tr}_b[\hat{G}_b\hat{B}(\theta,0)\hat{\rho}_a\otimes\left(|0\rangle_b\langle0|\right)\hat{B}^\dag(\theta,0)]
\approx \tan^2{\theta\over 2}\hat{a} ~t^{\hat{a}^\dag\hat{a}}\hat{\rho}_a t^{\hat{a}^\dag\hat{a}}~\hat{a}^\dag.
\label{sub-on-off}
\end{equation}
Like in Eq.~(\ref{sub-bs-ineff-s}) for the consideration of photodetector inefficiency, the use of an on-off detector inevitably results in the conditioned state to be mixed but under the usual condition, the resultant state is nearly a single-photon annihilated state.

We note that all the schemes to add and subtract a photon are conditional based on projective measurements.  A photon is added and subtracted conditioned on the initial preparation and final measurement outcome. In order to simulate annihilation or creation, we have found that to put or to take away extra unit of energy is not enough but the interaction with the process has to be weak.  Otherwise, the field state gets extra kick due to involvement of higher-order terms in the Taylor expansion of the interaction operator.

\subsection{Quantum scissors}
\label{q-sci}
Before we finish this section, we would like to briefly review a quantum engineering scheme based on the projective measurement like photon annihilation and creation introduced above. Earlier, Pegg, Phillips and Barnett~\cite{pegg-barnett} pioneered a scheme called the quantum scissors for state truncation.  The quantum scissors are composed of two concatenated beam splitters and photodetectors.  The output mode $b$ of the first beam splitter is arranged to be an input to the second beam splitter.  A single photon is input to mode $a$ of the first beam splitter while no photon is input to mode $b$ and a field of state $|\psi_{sc}\rangle$ to input mode $c$, where $c$ is to denote an input mode of the second beam splitter.  If the state is $|\psi_{sc}\rangle=\gamma_0|0\rangle+ \gamma_1|1\rangle + \gamma_2|2\rangle+ \cdots$ then the final state at the three outputs is
\begin{eqnarray}
{\gamma_0 \over \sqrt{2}}|1\rangle_a|0\rangle_b|0\rangle_c - {\gamma_0 \over 2}|0\rangle_a(|1\rangle_b|0\rangle_c &-& |0\rangle_b|1\rangle_c)
+ {\gamma_1 \over 2}|1\rangle_a(|0\rangle_b|1\rangle_c - |1\rangle_b|0\rangle_c)\nonumber \\
&&- {\gamma_1 \over 2}|0\rangle_a(|0\rangle_b|2\rangle_c - |2\rangle_b|0\rangle_c)+\cdots.
\label{scissors}
\end{eqnarray}
Now, conditioned on the measurement outcome 0 and 1 at modes $b$ and $c$ respectively, we effectively select the state in mode $a$, as
\begin{equation}
{\cal N}\left({\gamma_0\over 2}|0\rangle_a+ {\gamma_1\over 2}|1\rangle_a\right)
\label{nn}
\end{equation}
where ${\cal N}$ is the normalisation factor.  We see that the input $|1\rangle_a$ has turned into a superposition state whose composition depends on the input state in mode $c$.  In fact, taking $|\psi_{sc}\rangle$ as a coherent state we have a fair amount of freedom to choose $\gamma_0$ and $\gamma_1$.

\section{Quantum-state engineering by photon addition and subtraction}
\label{add-dak}
Dakna and coworkers~\cite{dakna99} proposed a scheme to generate a travelling field of an arbitrary superposition of Fock states.  Their scheme is based on the sequence of displacement operation $\hat{D}(\alpha)$ and addition of a single photon.  As a simple example, let us consider to generate a superposition of $|0\rangle$ and $|1\rangle$, which may be compared with a quantum scissors scheme to generate such the superposition as in Section~\ref{q-sci}.  The arbitrary superposition state $c_0|0\rangle + c_1|1\rangle$ with arbitrary complex numbers $c_0$ and $c_1$, can be written as
\begin{equation}
c_1\left({c_0\over c_1}|0\rangle+|1\rangle\right).
\label{arb-zero-one}
\end{equation}
Using the creation and displacement operators, state
(\ref{arb-zero-one}) can be written as
\begin{equation}
c_1\left(\hat{a}^\dag+{c_0\over c_1}\right)|0\rangle = c_1
\hat{D}^\dag\left({c_0\over
c_1}\right)\hat{a}^\dag\hat{D}\left({c_0\over c_1}\right)|0\rangle
\label{q-eng-add}
\end{equation}
where Eq.~(\ref{displace-ope}) has been used.  Thus, sending a vacuum to a sequence of a displacement operation, single-photon addition and another displacement operation, we obtain the state (\ref{arb-zero-one}).

\begin{figure}[b]
\centering{\includegraphics{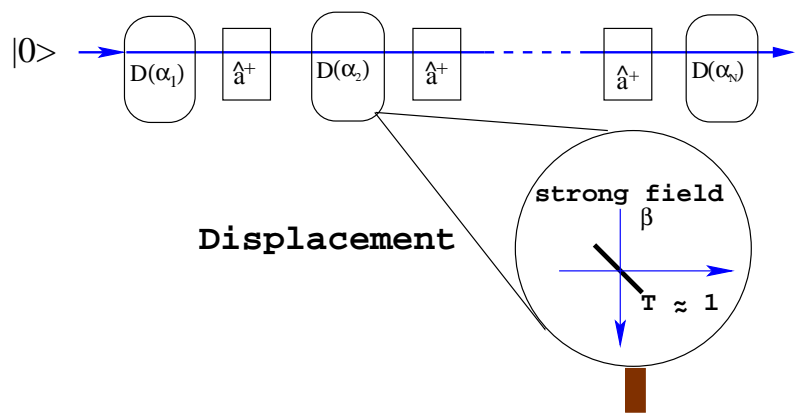}}
\caption{Diagrammatic sketch of the scheme to  generate an
arbitrary superposition of Fock states $\sum_{n=0}^N c_n|n\rangle$.
$\hat{a}^\dag$ denotes the operation of a single-photon addition and
$\hat{D}(\alpha_i)$ ($i=1,...,N$) is the displacement operation,
which can be realised by the use of a beam splitter of high
transmittivity $t$ and a strong coherent field of amplitude
$\beta$ as shown in the magnifier.  The displacement $\alpha=\beta\sqrt{1-T}$ with $T=|t|^2$} \label{fig:5}
\end{figure}

Using the equivalence of any superposition state
\begin{equation}
|\Psi\rangle=\sum_{n=0}^N c_n|n\rangle=\sum_{n=0}^N{c_n\over\sqrt{n!}}(\hat{a}^\dag)^n|0\rangle
\label{any-supp}
\end{equation}
to
\begin{equation}
|\Psi\rangle=\Pi_{n=1}^N(\hat{a}^\dag+\alpha_n^*)|0\rangle= \Pi_{n=1}^N\hat{D}^\dag(\alpha_n)\hat{a}^\dag\hat{D}(\alpha_n)|0\rangle
\label{any-supp-2}
\end{equation}
Dakna and coworkers~\cite{dakna99} showed that an arbitrary superposition of $|0\rangle, |1\rangle,\cdots, |N\rangle$ can be generated by a concatenation of $N$ units of operations composed of displacement, single-photon addition and another displacement as shown in Fig.~\ref{fig:5}.

The displacement of a quantum state in phase space can be
realised~\cite{paris} using a beam splitter of high transmittivity
$t\approx 1$ and a strong coherent field of its amplitude $\beta$.
Let us consider a beam splitter operation where one input mode is a
coherent field of $|\beta\rangle_b$:
\begin{equation}
\hat{B}(\theta,\phi=0)|\beta\rangle_b=\mbox{e}^{{\theta\over
2}(\hat{a}^\dag\hat{b}-\hat{a}\hat{b}^\dag)} |\beta\rangle_b
\sim\mbox{e}^{{\theta\over
2}(\hat{a}^\dag\beta-\hat{a}\beta^*)}|\beta\rangle_b \label{approx0}
\end{equation}
where the approximation was done for the condition of a large
amplitude $\beta$.  The result of the operation is the displacement
of ${\theta\over 2}\beta$ for mode $a$.  More rigorously, we can
discuss the displacement operation as follows.  Any field state of
density operator $\hat{\rho}$ can be written as a weighted sum of
coherent states as in Eq.~(\ref{p-function}).  Thus by beam
splitting a field of an arbitrary state in mode $a$ and a coherent
field $|\beta\rangle$ in mode $b$, we have
\begin{eqnarray}
\hat{\rho}_{ab}&=&\hat{B}(\theta,0)\hat{\rho}_a\otimes|\beta\rangle_b\langle\beta|\hat{B}^\dag(\theta,0) \nonumber\\
&=& \hat{B}(\theta,0)\int d^2\alpha P(\alpha)|\alpha\rangle_a\langle\alpha| \otimes |\beta\rangle_b\langle\beta|\hat{B}^\dag(\theta,0)
\nonumber \\
&=&\int d^2\alpha P(\alpha)|t\alpha + r\beta\rangle_a\langle
t\alpha+r\beta|\otimes |t\beta-r\alpha\rangle_b\langle
t\beta-r\alpha|.
\end{eqnarray}
The state of mode $a$ regardless of mode $b$ can be obtained by tracing over mode $b$:
\begin{equation}
\mbox{Tr}_b\hat{\rho}_{ab}=\int d^2\alpha P(\alpha)|t\alpha+r\beta\rangle_a\langle t\alpha+r\beta|
\label{approx1}
\end{equation}
which is approximated to the displaced state
\begin{equation}
\hat{D}(r\beta)\hat{\rho}_a\hat{D}^\dag(r\beta)=\int d^2\alpha P(\alpha) |\alpha+r\beta\rangle_a\langle\alpha+r\beta|
\label{approx2}
\end{equation}
when $t\approx 1$.  As $r=\sin{\theta\over 2}\approx {\theta\over 2}$ for $\theta\approx 0$, Eqs.~(\ref{approx0}) and (\ref{approx2}) are in good agreement.  However, as can be seen in Eq.~(\ref{approx1}), the displacement operation is never exact as when $t=1$, $r=0$ and a state is never displaced and when $t<1$, state (\ref{approx1}) is never the same as (\ref{approx2}).

While adding a photon is difficult to realise, subtraction is relatively easy (an experiment of single-photon addition will be explained later in this paper).  Fiur\'{a}\v{s}ek, Garc\'{i}a-Patr\'{o}n and Cerf~\cite{fiu1} proposed to engineer a quantum state by a sequence of subtraction and displacement of a squeezed vacuum.  First, a squeezed vacuum is displaced before a single photon is subtracted from it.  After another displacement operation, the state is antisqueezed.  All these processes are summarised as follows:
\begin{equation}
|\Psi\rangle_{out}=\hat{S}_{1}^\dag(\zeta_2)\hat{D}(\alpha_2)\hat{a}\hat{D}(\alpha_1)\hat{S}(\zeta_1) |0\rangle.
\label{q-eng-sub-0}
\end{equation}
If $\alpha_1=\alpha_2$ and $\zeta_1=\zeta_2$ then
\begin{equation}
|\Psi\rangle_{out}=
\hat{S}_{1}^\dag(\zeta_1)(\hat{a}+\alpha_1)\hat{S}(\zeta_1)
|0\rangle= -\sinh\zeta_1|1\rangle+\alpha_1|0\rangle,
\label{q-eng-sub}
\end{equation}
where the unitary transformation (\ref{unitary-s1}) has been used.
By an appropriate choice of $\zeta_1$ and $\alpha_1$, we can have an
arbitrary superposition of $|0\rangle$ and $|1\rangle$ states. It
can be proven~\cite{fiu1} that we can achieve an arbitrary
superposition of Fock states, by concatenation of the units composed
of subtraction and displacement of the squeezed vacuum together with
the final antisqueezing.  The proof follows a similar analysis shown
earlier in this Section. In the next subsection, we will show how
to produce a specific quantum state using a squeezed vacuum.

\subsection{Production of a cat state}
The sequence of photon subtraction or addition is a demanding task in experiments.
As a simple example of quantum-state engineering, let us consider recent experimental triumphs to produce linear superpositions of coherent states, by subtracting a photon from a squeezed state.
It has been a long  awaited dream to produce a coherent
superposition state (\ref{superposition}), which was briefly
discussed in Section \ref{section:2}.  A coherent state is
considered to be at the boundary of classical and nonclassical
worlds and its superposition is accepted to show the quantum nature
of Schr\"{o}dinger's cat paradox\footnote{In his argument
Schr\"{o}dinger tried to show how the microscopic world which may be
under the laws of quantum mechanics is directly coupled to the
macroscopic world which is under the laws of conventional physics at
the time.  In this paradox, a cat is placed in a steel box and a
poisonous gas is set to be released when an atom, connected to the
container of the gas, decays. Reading \cite{wheeler}, it is not
clear that the state described in the paradox is an entangled state
between the atomic state and the destiny of the cat or it is a
simple superposition of the cat's destiny.  Had it been the former,
the state for the cat and atom will be
$|no-decay\rangle|alive\rangle+ |decay\rangle|dead\rangle$ assuming
the equal amplitude of the two events of atomic decay and non-decay.
Here, normalisation has been neglected.  If the paradox is simply
described by the superposition of a cat state, then
$|dead\rangle+|alive\rangle$ will describe the paradox.}.  There has
been a difficulty in producing such the superposition state in a
travelling field because of the unavailability of extremely high
nonlinearity required~\cite{jeong-kim}.  The coherent state is a
Poissonian weighted superposition of Fock states~\cite{glauber03}
\begin{equation}
|\beta\rangle=\mbox{e}^{-|\beta|^2/2}\sum_{n=0}^\infty{\beta^n\over\sqrt{n!}}|n\rangle
\end{equation}
thus assuming the amplitude $\beta$ to be real, the coherent superposition state (\ref{superposition}) is
\begin{equation}
|\Psi_e\rangle={1\over\sqrt{2\cosh\beta^2}}\sum_{n=0}^\infty
\frac{\beta^{2n}}{\sqrt{(2n)!}}|2n\rangle \label{even-co}
\end{equation}
for the phase $\phi_{sch}=0$, and
\begin{equation}
|\Psi_o\rangle={1\over\sqrt{2\sinh\beta^2}}\sum_{n=0}^\infty
\frac{\beta^{2n+1}}{\sqrt{(2n+1)!}}|2n+1\rangle \label{odd-co}
\end{equation}
for $\phi_{sch}=\pi$.  It is clear that $|\Psi_e\rangle$
($|\Psi_o\rangle$)  has the non-zero probabilities only for even (odd) numbers of photons hence the name the even (odd)
coherent state. The Poissonian weight is shown in
Fig.~\ref{fig:coh-photon number}.
\begin{figure}[b]
\centering{\includegraphics{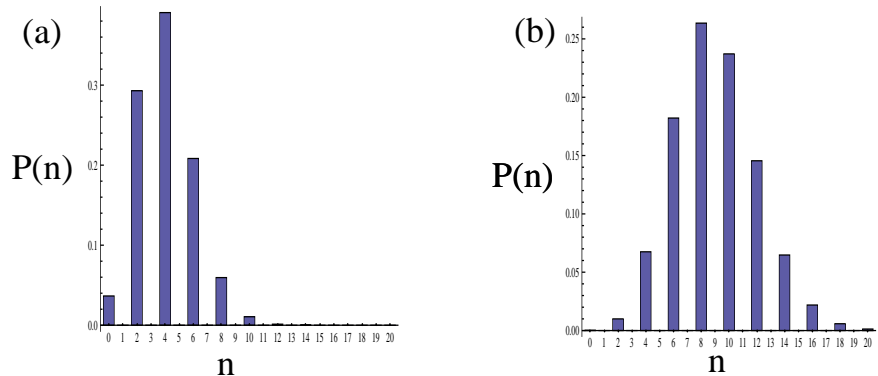}}
 \caption{Photon number distributions for the even coherent states of
{\bf(a)}: $\beta=2$ {\bf (b)}: $\beta=3$.} \label{fig:coh-photon
number}
\end{figure}

Recalling  that the squeezed vacuum has non-zero probabilities of having only the even numbers of photons, as shown in Eq.~(\ref{sq1-state}), we can immediately recognise some analogy between the photon number distributions of the coherent
superposition state and the squeezed vacuum\footnote{We
analyse the similarity of two states based only on their photon
number distribution in this part of the paper.  This analysis is
valid when the two states are pure and the weights of Fock-state
components are real and positive.  Although it is dangerous to identify a quantum state under such the restrictive constraints, it provides an intuitive picture so to use it in this part of the paper.}.  While the
weight function in Eq.~(\ref{sq1-state}) is a decreasing function
with the peak at $n=0$, that in Eq.~(\ref{even-co}) is Poissonian
with the peak at $\beta^2$.  If we can push the peak of the weight
function (\ref{sq1-state}), it seems to be possible to get the two distributions
closer to each other. By annihilating a photon from the squeezed
vacuum, the state becomes (without normalisation)
\begin{equation}
\hat{a}\hat{S}|0\rangle=\hat{S}\hat{S}^\dag\hat{a}\hat{S}|0\rangle=\hat{S}(\hat{a}\cosh\zeta +\hat{a}^\dag\sinh\zeta)|0\rangle\rightarrow\hat{S}\hat{a}^\dag|0\rangle
\label{ann-sq}
\end{equation}
where the unitarity of the squeezing operator has been used together
with the transformation properties (\ref{unitary-s1}).  The argument
$\zeta$ of the squeezing operator has been omitted to make the
notation simple and the sign $\rightarrow$ means that its left hand
side becomes the same as the right hand side after
normalisation.  We note that annihilating a photon from  the
squeezed vacuum gives the same result as squeezing a single photon
state.  Furthermore, we will see that annihilating a photon in the
squeezed vacuum gives the same impact as adding a photon to the
squeezed vacuum.  By applying a creation operator to the squeezed
vacuum, we get
\begin{equation}
\hat{a}^\dag\hat{S}|0\rangle=\hat{S}\hat{S}^\dag\hat{a}^\dag\hat{S}|0\rangle= \hat{S}(\hat{a}^\dag\cosh\zeta +\hat{a}\sinh\zeta)|0\rangle\rightarrow\hat{S}\hat{a}^\dag|0\rangle.
\label{add-sq}
\end{equation}

It is  surprising that adding a single photon to a squeezed vacuum
gives the same impact as annihilating a single photon from it.  A
starting point to understand this is the fact that there are only
even numbers of photons in the squeezed vacuum.  By subtracting a
photon, we know that there were at least two photons in the field, as otherwise it would not have been possible to subtract any photon.
By subtracting a photon, we now know that there is at least one
photon in the field.  Thus, subtracting a photon, we annihilate a
chance of having no photons and results in the same effect as adding
a photon to the field.  Here, we remind ourselves that the process is conditional.

\begin{figure}[b]
\centering{\includegraphics{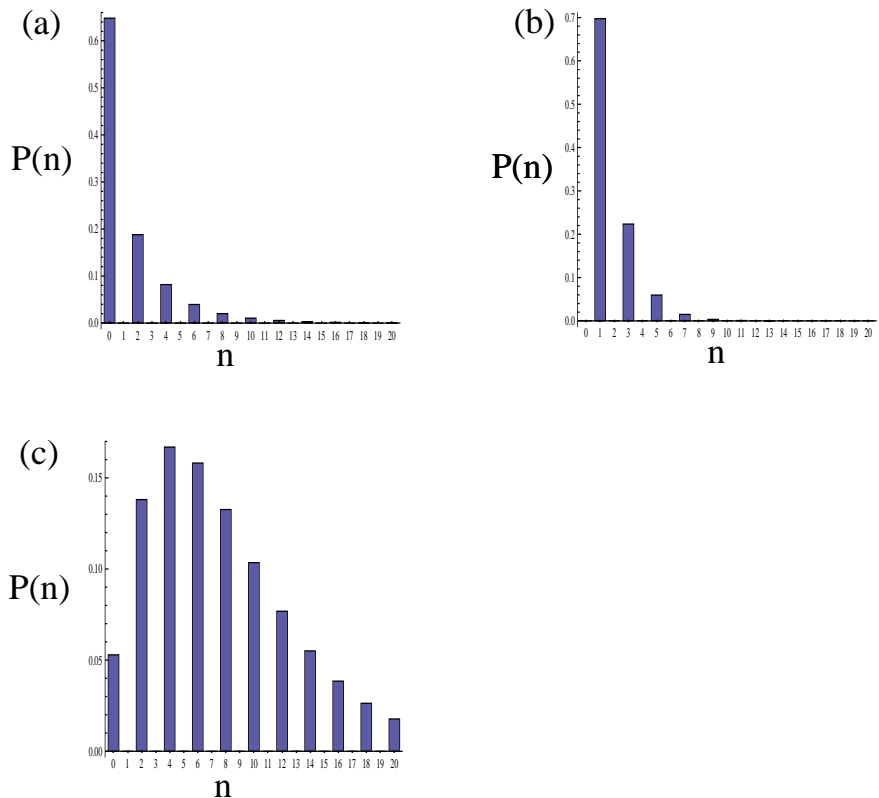}}
\caption{Photon number distributions for {\bf(a)}: the squeezed
vacuum {\bf (b)}: the single-photon annihilated squeezed vacuum, and
{\bf(c)}: the two-photon annihilated squeezed vacuum.  The squeezing
parameters are $\zeta=1$ for {\bf (a)} and {\bf (c)} and $\zeta=0.5$ for {\bf (b)}. }
\label{fig:sq-photon number}
\end{figure}

By subtracting a photon, we have been able to move the peak from
zero photon to nonzero photons.  This is well seen in
Fig.~\ref{fig:sq-photon number}.  The position of the peak depends
on how many photons are annihilated and how much the state is
squeezed initially.  The reason why the peak moves to a higher
photon number is because of the coefficient $\sqrt{n}$ of the annihilation operation as
shown in Eq.~(\ref{1}).  This coefficient gives a higher weight to the initial higher-number photon component.

For the photon subtracted squeezed vacuum to simulate  the coherent
superposition state, the weight distribution of its component states
should be close to that of the coherent superposition state.  The higher the squeezing, the flatter the
weight distribution for the initial squeezed vacuum.  Thus by subtracting a photon we
have a longer and thicker tail in the weight distribution than in
the Poissonian distribution for the coherent superposition state.
By an appropriate choice of the squeezing parameter, the
single-photon annihilated squeezed state is observed to simulate the
coherent superposition state~\cite{dakna97}. A sequence of single-photon
annihilations will gradually increase the
weights of higher numbers of photons (Noting that a
coherent state is an eigenstate of an annihilation operator, by
annihilating a photon from an even coherent state, it turns into an
odd coherent superposition state and vice versa~\cite{knight-barry}. However, the
photon-subtracted squeezed vacuum changes its nature slightly as it
is not a perfect coherent superposition state.  As the number of subtraction operations grows the resultant state digresses from the coherent superposition state.).  Now, we face a problem
to measure how close two states are to each other.  One widely
accepted measure is the fidelity which is simply an overlap
$|\langle\psi|\phi\rangle|^2$ between two states $|\psi\rangle$ and
$|\phi\rangle$.  This measure can be used if at least one of the
fields is in a pure state.  If the other state is mixed so to be
represented only by its density operator $\hat{\rho}_\phi$ then the
fidelity is
\begin{equation}
{\cal F}=\langle\psi|\hat{\rho}_\phi|\psi\rangle
\label{fidelity-pure}
\end{equation}
which can be calculated using the overlap between their Wigner functions and Weyl  characteristic functions:
\begin{equation}
{\cal F}=\pi\int d^2\alpha W_\psi(\alpha)W_\phi(\alpha)= {1\over\pi}\int d^2\xi C_\psi(\xi) C_\phi(-\xi)
\label{fidelity-wigner}
\end{equation}
When the two states are the same, the
fidelity is 1 with the perfect overlap while when the two
states are orthogonal to each other, it is 0.  The fidelity, however, is not
an ultimate measure as it can smear out some useful information.
Moreover, in general, we do not know how to measure the fidelity
experimentally, except a few cases~\cite{kim-munro}.  For a physical
system which is described in a small Hilbert space, the distance
measure can be more useful and it has its implication in
experimental realisation~\cite{kim-lee}.

It was calculated theoretically~\cite{kim-jeong-park} that the
fidelity between the coherent superposition state and the
single-photon annihilated squeezed vacuum is found to be as high as
0.99 when the squeezing parameter $\zeta=0.43$  and the coherent
state amplitude $\beta=1.16$.  The production of such the
superposition state was successfully observed experimentally
~\cite{grangier02}, under the name of a {\it Schr\"{o}dinger kitten
state} (While Ref.~\cite{grangier02}
reported the production of the coherent superposition state in a
pulsed field, Ref.~\cite{polzik01} demonstrated the production of
such the state in a continuous wave~\cite{molmer02}.) because the amplitude of the component state is very small, which is
a problem to use it for many of the tasks in quantum information
processing. In their work,
Ourjoumtsev and coworkers~\cite{grangier03} have used a somewhat
different scheme to realise a coherent superposition state with a
larger amplitude as explained below.

A squeezed superposition of Fock states may be useful for various
applications.  Going back to the Fiur\'{a}\v{s}ek and coworker's
scheme~\cite{fiu1},  a squeezed superposition state is generated by omitting the final squeezing.  In this respect, the generation of a squeezed Schr\"{o}dinger cat state has recently been demonstrated
experimentally~\cite{grangier03}.  Here, the superposition of two
squeezed coherent states with amplitude $\alpha=\pm\sqrt{2.6}$ and
squeezing of 3.5dB, was produced. The scheme starts from the
generation of a Fock state $|n\rangle$.  The Fock state is sent to a
50:50 beam splitter. Then a homodyne detector is placed at one
output port in order to select a state at the other output, conditioned on
the homodyne measurement outcome $q=0$.

The state conditionally generated is
\begin{equation}
|\Psi_{sc}\rangle_a=~_b\langle q(0)|\hat{B}\left(\theta=-{\pi\over 2}, \phi=0\right)|n\rangle_a|0\rangle_b
\label{sq-coh-1}
\end{equation}
where the eigenstate $|q(x)\rangle$ of the quadrature operator $\hat{q}$ can be written as the superposition of Fock states
\begin{equation}
|q(x)\rangle={1\over \pi^{1/4}}\sum_{n=0}^\infty {1\over\sqrt{2^nn!}}\mbox{e}^{-x^2/2}H_n(x)|n\rangle.
\label{q-x}
\end{equation}
A Hermite polynomial $H_n(x)$ for argument 0 has a non-zero value only for an even index, {\it i.e.}, $H_{2n}(0)=(-2)^n(2n-1)!!$ and $H_{2n+1}=0$.  Thus
\begin{equation}
|q(0)\rangle={1\over \pi^{1/4}}\sum_{n=0}^\infty (-1)^n
\frac{(2n-1)!!}{\sqrt{(2n)!}}|2n\rangle \label{q-x-0}
\end{equation}
Using the decomposition of the beam splitter operator (\ref{beam-splitter-decom}), the beam splitter operation in Eq.~(\ref{sq-coh-1}) is
\begin{eqnarray}
\hat{B}(\theta=\pi/2 &,& \phi=0)|n\rangle_a|0\rangle_b=\left({1\over\sqrt{2}}\right)^n\mbox{e}^{\hat{a}\hat{b}^\dag} |n\rangle_a|0\rangle_b
\nonumber \\
&&= \left({1\over\sqrt{2}}\right)^n \sum_{k=0}^n\left(\begin{array}{c}
n \cr
k\cr
\end{array}\right) |n-k\rangle_a|k\rangle_b.
\label{b-0}
\end{eqnarray}

Conditioned on $q=0$ for mode $b$, we see that mode $a$ has only even or odd numbers of photons depending on the initial preparation of the Fock state.  For example, if $n$ is even, the state generated is of even numbers of photons.  The weight function of the Fock-state components is determined by the binomial distribution and the Hermite polynomial.  As the peak is shifted by the original excitation of the Fock state, we can see a possibility of generating a cat state with a larger amplitude.  The squeezing of the cat state appears due to the final weight function as the overlap of the Hermite polynomial and the binomial distribution.  We note that under this scheme,  the coherent superposition state, which lies in the infinite-dimensional Hilbert space, is realised in a $n$-dimensional Hilbert space restricted by the initial preparation of the Fock state.

\begin{figure}[b]
\centering{\includegraphics{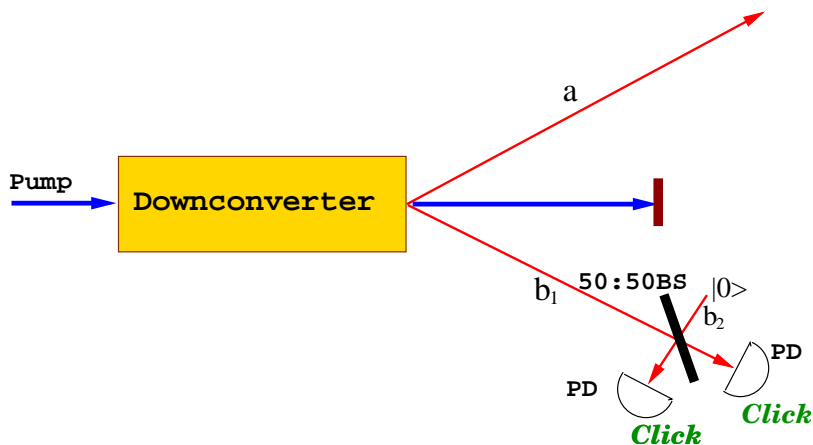}}
\caption{An optical downconverter produces twin photons into modes $a$ and $b_1$.  Conditioned on the projection of mode $b_1$ onto $|2\rangle$, we generate the two-photon state $|2\rangle$ in mode $a$.  The condition is based on counting single photons ({\it click}s) in modes $b_1$ and $b_2$ at photodetectors (PDs). }
\label{fig:6}
\end{figure}

In the experimental realisation, the bottleneck of this scheme is the initial preparation of the
Fock state.  In 2006, the experimental group in
Paris~\cite{grangier04} successfully produced a two-photon Fock
state $|2\rangle$ using an optical parametric amplifier.  In this
experiment, the non-degenerate downconversion converts one pump photon to two photons into two different modes as
explained in the two-mode squeezing operator in Eq.~(\ref{sq2}).  At
one output mode of the optical parametric amplifier, a 50:50 beam
splitter is placed as shown in Fig.~\ref{fig:6}.  Detecting a photon
at each of the output mode the input field is projected onto
\begin{equation}
\underbrace{~_{b1}\langle 1|_{b2}\langle 1 |\hat{B}\left(-{\pi\over
2},0\right)}_{\mbox{output}} \Longrightarrow
\underbrace{{1\over\sqrt{2}}(~_{b1}\langle 2|_{b2}\langle 0|+
_{b1}\langle 0|_{b2}\langle 2|)}_{\mbox{input}} \label{2-proj}
\end{equation}
As input mode $b2$ is served by the vacuum, the coincidence count is to project the $b1$ mode onto the two-photon Fock state. Due to the twin-photon nature of the downconversion, when the field mode $b$ is projected onto the two-photon Fock state, the field mode $a$ is also in the two-photon Fock state.  In this way, the experimental group in Paris produced $|2\rangle$ state and proved it by the full reconstruction of its Wigner function.

\subsection{Photon addition using parametric down converter}
We have considered schemes to add a photon using a cavity or a beam splitter in Section \ref{add-subtract}.  In this subsection, we show another way to add a photon to a travelling field which has recently been realised experimentally.
Agarwal and Tara~\cite{agarwal-tara} found that adding a definite
number of photons to a coherent state, the state turns into a highly
nonclassical state, showing a sub-Poissonian character and quadrature
squeezing.  Lee~\cite{lee} later realised that if a single photon is
added to any field state (pure or mixed), the state turns into a
nonclassical state for which a positive well-behaved $P$ function
cannot be imposed. He defines the depth of nonclassicality using the
properties of the quasiprobability functions.  As shown in
Eq.~(\ref{s-s}), the quasiprobability functions are connected to
each other.  At one end of the spectrum ($s=1$), the $P$ function
may not even exist as a rational function for a nonclassical field
while at the other end of the spectrum ($s=-1$), the $Q$ function is
always positive and possesses all the properties as a proper
probability distribution.  When $s=1$, Eq.~(\ref{s-s}) becomes
\begin{equation}
P(\alpha,\tau)=\frac{1}{\pi\tau}\int_{-\infty}^{\infty}d^2\beta \exp\left( \frac{-|\alpha-\beta|^2}{\tau} \right)P(\beta)~~~;~~~\tau=\frac{1-s^\prime}{2}.
\label{noncl-depth}
\end{equation}
Lee defined the depth of nonclassicality based on how well  the
quasiprobability function behaves like a proper probability
distribution. The minimum $\tau$ which brings $P(\alpha,\tau)$ to a
proper probability function is the depth of nonclassicality. This
means that through the spectrum of $s$, if a state has its
quasiprobability function $P(\alpha,s)$ positive well-behaved only
for its $Q$ function, the state is said to be most nonclassical.
This definition is related to the resilience of the nonclassicality
to the vacuum noise as its influence appears in the form of Gaussian
convolution (See \cite{kim-imoto}).  Lee proves that for any state
of density operator $\hat{\rho}=\sum_{n,m=0}^\infty
\rho(n,m)|n\rangle\langle m|$, if $\rho(0,0)=0$ then the
nonclassical depth of the state is 1. According to this theorem,
even a very chaotic thermal field at a very high temperature can
become nonclassical only by adding a single photon to it.  Of
course, how useful it can be for any kind of quantum processing is
another matter.

\begin{figure}[b]
\centering{\includegraphics{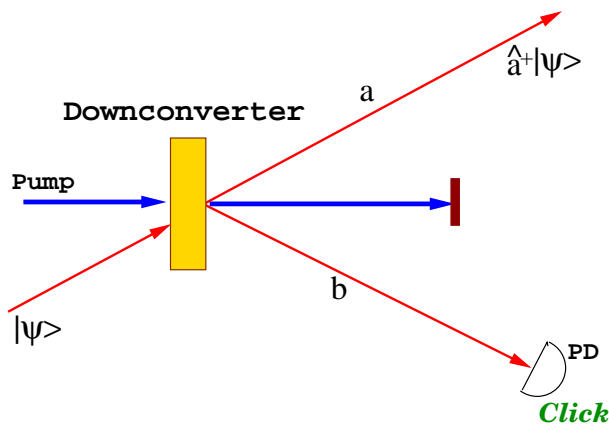}}
\caption{Zavatta, Viciani and Bellini~\cite{bellini01} successfully added a photon into a coherent field using a downconverter.  Conditioned on photodetection in mode $b$ a photon is added to the input field $|\psi\rangle$. }
\label{fig:7}
\end{figure}

Even though the addition of a single photon  (or a definite number
of photons) guarantees nonclassical behaviour, it had to wait for
long before such the operation was first realised experimentally by
Zavatta, Viciani and Bellini~\cite{bellini01}.  Realising that
adding a photon conditioned on {\it `no' detection} of a photon
using a beam splitter and a single photon input is experimentally
demanding, they add a photon using a parametric downconversion
process.  If a photon is detected in mode $b$ (see
Fig.~\ref{fig:7}), we know that its twin photon is in mode $a$. If
$|\psi\rangle$ is an input to mode $a$, the output field conditioned
on a photodetection in mode $b$ is ${\cal
N}\hat{a}^\dag|\psi\rangle$ with normalisation ${\cal N}$.  By
adding a photon to an input coherent state, Zavatta and coworkers
claim the observation of the transition from a classical to a
quantum state.  They show this by the full reconstruction of the
Wigner functions.  They also show the nonclassicality of a
photon-added thermal state in another experiment~\cite{bellini02}.

The parametric downconversion is described by the two-mode squeezing operator (\ref{sq2}), which can be decomposed into (See Appendix 5 in Ref.~\cite{barnett-radmore})
\begin{equation}
\hat{S}_2(\zeta)=\mbox{e}^{-\hat{a}^\dag\hat{b}^\dag\tanh\zeta}
\mbox{e}^{-(\hat{a}^\dag\hat{a}+
\hat{b}\hat{b}^\dag)\ln(\cosh\zeta)}\mbox{e}^{\hat{b}\hat{a}\tanh\zeta}.
\label{s2-decom}
\end{equation}
Detecting a photon in mode $b$ at the output while its input is vacuum, is equivalent to the addition of a photon in mode $a$:
\begin{equation}
~_b\langle 1|\hat{S}_2(\zeta)|0\rangle_b=-
\frac{\sinh\zeta}{\cosh^2\zeta}\hat{a}^\dag
(\cosh\zeta)^{-\hat{a}^\dag\hat{a}} \label{s2-add}
\end{equation}
where the constant term should  be ignored as it will disappear
during the normalisation.  When the squeezing is small, in other
words, when the efficiency of the parametric downconversion is
small, and the input photon number is not large, the term
$(\cosh\zeta)^{-\hat{a}^\dag\hat{a}}$ is negligible.  Thus the
process becomes to simulate the photon creation operation
$\hat{a}^\dag$.

\subsection{A sequence of photon addition and subtraction}
Photon addition and subtraction are now experimentally plausible
schemes and in this section, we combine the two schemes.  Quantum
mechanically, photon addition and subtraction correspond to bosonic
creation and annihilation operations, respectively.  As the two
operators do not commute, the two different sequences of addition
and subtraction will bear different states.  Here, we show how
significant the differences are.

By annihilating and then creating a photon to a coherent state
$|\beta\rangle$, the state becomes
 \begin{equation}
|\psi_c^{as}\rangle=\sqrt{ {\cal N}_c^{as}}\hat{a}^\dag\hat{a}|\beta\rangle~~~;~~~{\cal N}_c^{as}=1/\bar{n}_c(1+\bar{n}_c)
\label{coherent-as}
\end{equation}
where $\bar{n}_c=|\beta|^2$ is the mean photon number of the
coherent field.  The photon number distribution for the resultant
state is
\begin{equation}
P_c^{as}(n)={\cal N}_c^{as}nP_c(n)~~~;~~~ P_c(n)=\mbox{e}^{-\bar{n}_c}\frac{\bar{n}_c^n}{n!},
\label{coherent-sa-pnd}
\end{equation}
which obviously has null zero-photon probability.  In order to
quantify the uncertainty of the photon number in comparison to the
Poissonian level, we define the $Q$-parameter~\cite{mandel},
$Q=(\Delta n)^2-\bar{n}$.  For $Q<1$, the field is said to manifest
sub-Poissonian statistics while for $Q>1$, the field is
super-Poissonian.  Considering the fact that a coherent state is an
eigenstate of $\hat{a}$, i.e.
$\hat{a}|\beta\rangle=\beta|\beta\rangle$ and the photon number
operator $\hat{n}=\hat{a}^\dag\hat{a}=\hat{a}\hat{a}^\dag-1$, we
find the $Q$-parameter for the field (\ref{coherent-as}) as
$Q=-1-(\bar{n}/(\bar{n}+1))^2$.  The field (\ref{coherent-as}) thus
is sub-Poissonian.

If we add a photon before subtracting one for an initial coherent state, the state is
\begin{equation}
|\psi_c^{sa}\rangle=\sqrt{ {\cal N}_c^{sa}}\hat{a}\hat{a}^\dag|\beta\rangle~~~;~~~{\cal N}_c^{sa}=1/(\bar{n}_c^2+3\bar{n}_c+1).
\label{coherent-sa}
\end{equation}
Its photon number distribution is
$P_c^{sa}(n)=(n+1)^2P_c(n)/(\bar{n}^2+3\bar{n}+1)$, which is
different from the initial $P_c(n)$.  We find that the $Q$ parameter
is negative for any value of $\beta$.

By subtracting then adding a photon to a thermal field, the density
operator
$\hat{\rho}_B=\sum\bar{n}_B^n/(1+\bar{n}_B)^{n+1}|n\rangle\langle
n|$ of the initial field becomes
\begin{equation}
\hat{\rho}_B^{as}={\cal
N}_{B}^{as}\hat{a}^\dag\hat{a}\hat{\rho}_B\hat{a}^\dag\hat{a}~~;~~
{\cal N}_B^{as}=1/(\bar{n}_B(2\bar{n}_B+1)). \label{thermal-as}
\end{equation}
The average photon number and the variance for the resultant state
are
\begin{equation}
\bar{n}_{B}^{as}=\frac{2\bar{n}_B(3\bar{n}_B+2)}{2\bar{n}_B+1}~~;~~(\Delta n_B^{as})^2=12\bar{n}_B^2+12\bar{n}_B+1,
\label{avg-thermal-as}
\end{equation}
which leads the $Q$ parameter smaller than zero when the average
photon number $\bar{n}_B$ of the initial thermal state is less than
around 0.6.

The fact that subtracting and adding a photon brings a thermal field to a sub-Poissonian state, may inspire us to conjecture that subtracting a photon can narrow down the width of the photon number distribution. By subtracting $\ell$ number of photons from the thermal field, the $Q$ parameter becomes $\bar{n}^2(\ell+1)$ which is always positive.  In fact, not the subtracting but the process of adding a photon turns the thermal field into a sub-Poissonian field as the resultant state has the $Q$ parameter equals $2\bar{n}^2-1$.

For the reverse process of subtracting after adding a photon, the state becomes
\begin{equation}
\hat{\rho}_B^{sa}={\cal N}_{B}^{sa}\hat{a}\hat{a}^\dag
\hat{\rho}_B\hat{a}\hat{a}^\dag~~;~~{\cal
N}_B^{sa}=1/((\bar{n}+1)(2\bar{n}+1)). \label{thermal-sa}
\end{equation}
The  average photon number in this case is smaller than that for
$\hat{\rho}_B^{as}$ by one.  Even though the photon number
distribution is not the same as the original one, the resultant
state is always super-Poissonian with $Q>0$.

In order to see the  negativity of the Wigner function, we use a
{\it theorem}~\cite{barnett-n} that if the $Q$ function becomes zero
at any point of phase space, the Wigner function should have
negative values.   Using this, we can easily check if a state is
nonclassical because the $Q$ function, which is defined in
Eq.~(\ref{q-function}), is much easier to calculate than the Wigner
function.

For $|\psi_c^{as}\rangle$, the $Q$ function is found to be
\begin{equation}
Q_c^{as}(\alpha)={{\cal N}_c^{as} \over \pi}|\beta|^2|\alpha|^2\mbox{e}^{-|\beta-\alpha|^2}
\label{q-coherent-as}
\end{equation}
which is zero at the origin of phase space.  Thus, the Wigner
function should manifest negativity and the resultant state is
nonclassical.  In fact, we can easily prove that any field becomes
nonclassical by subtracting and adding a photon.  The density
operator of a classical state can be written as
$\hat{\rho}_{cl}=\int
P_{cl}(\beta)|\beta\rangle\langle\beta|d^2\beta$, where
$P_{cl}(\alpha)$ is a positive well-behaved function.    The
$Q$ function for the field after subtracting and adding a photon is
the weighted sum of (\ref{q-coherent-as}) with a proper
normalisation.  Thus the Wigner function should be negative at some
points of phase space as the $Q$ function is zero at the origin of
phase space.  This can be understood easily as the state after
photon subtraction and addition is  never empty and according to Lee~\cite{lee} this field of zero null-photon probability is
nonclassical.

For $|\psi_c^{sa}\rangle$, the $Q$ function is
\begin{equation}
Q_c^{sa}(\alpha)={{\cal N}_c^{sa} \over \pi}|\alpha^*\beta+1|^2\mbox{e}^{-|\beta-\alpha|^2}
\label{q-coherent-sa}
\end{equation}
which surely becomes zero when $\alpha^*\beta+1=0$ so the field is
nonclassical.  However, we cannot conclude that any classical state
becomes nonclassical by adding and then subtracting a photon.
It is because
\begin{equation}
Q_{cl}^{sa}(\alpha)\propto\int
d^2\beta|\alpha^*\beta+1|^2\mbox{e}^{-|\beta-\alpha|^2}P_{cl}(\beta)
\label{q-classical-sa}
\end{equation}
does not seem to be zero at any point of phase space.    Here, the
integrand is a positive function and $\alpha$ cannot make the
integrand zero regardless of $\beta$.  The Wigner function thus
does not become negative.  However we have not proven that the
initial classical state does not turn into a nonclassical state
after photon addition and subtraction.  Even though the negativity
in the Wigner function is a strong proof of a state being nonclassical,
the negativity or non-existence of the $P$ function is another
criterion for nonclassicality.

\begin{figure}
\centering{\includegraphics[width=.8\textwidth]{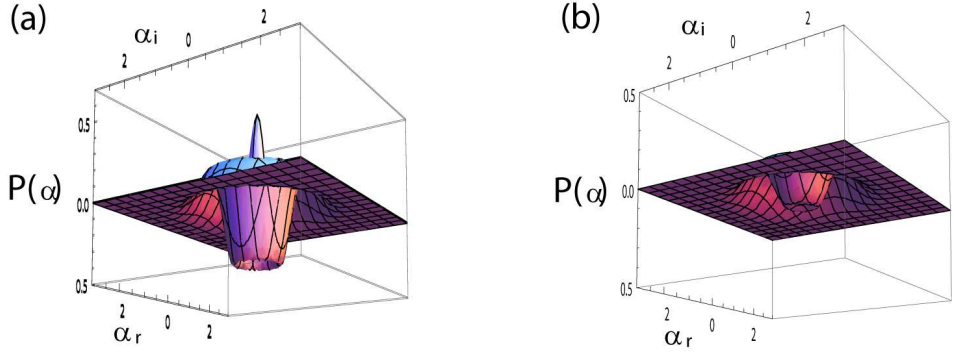}}
\caption{The $P$ functions for $\hat{\rho}_B^{as}$ which is obtained after subtracting then adding a photon to a thermal field {\bf (a)} and for $\hat{\rho}_B^{sa}$ which is obtained after adding then subtracting a photon for a thermal field {\bf (b)}.  The average photon number for the initial thermal field is $\bar{n}_B =0.5$. }
\label{fig:p-function}
\end{figure}

In order to find out the nonclassicality of a state after adding and
subtracting a photon for an initial classical state, we consider an
example of the thermal field.  After a straightforward calculation, we
find the $P$ function for $\hat{\rho}_B^{sa}$ in (\ref{thermal-sa}):
\begin{equation}
P_B^{sa}(\alpha)={\cal N}_{B}^{sa}(\bar{n}_B+1)|\alpha|^2\left[\frac{\bar{n}_B+1}{\bar{n}_B}|\alpha|^2-1\right]\mbox{e}^{-|\alpha|^2/\bar{n}_B},
\label{P-thermal-sa}
\end{equation}
which becomes negative when the term in bracket is negative.  Thus
the resultant state is nonclassical.  As a comparison, the $P$
function for  $\hat{\rho}_B^{as}$ in (\ref{thermal-sa}) is found to
be
\begin{equation}
P_{B}^{as}(\alpha)={\cal N}_{B}^{as}\left\{\left[\frac{\bar{n}_B+1}{\bar{n}_B}|\alpha|^2-1\right]^2 -\frac{\bar{n}_B+1}{\bar{n}_B}|\alpha|^2\right\}\mbox{e}^{-|\alpha|^2/\bar{n}_B}.
\label{P-thermal-as}
\end{equation}
The $P$ functions are plotted in Fig.~\ref{fig:p-function} for $\bar{n}_B=0.5$.

The quasiprobability functions have been considered for two
initial states.  Any sequence of photon addition and subtraction
brings the state into the nonclassical state in our studies. In fact, the sequence of
photon addition and subtraction has been experimentally realised for
the input thermal field~\cite{bellini-kim}. In this work, the
reconstructed Wigner function shows the negative dip for the photon
subtracted and added state.

\section{photon subtraction to increase entanglement}
\subsection{A brief review of nonlocality, entanglement and quantum teleportation}
When two systems are correlated, knowing one system reduces the uncertainty of the other system.  The correlation allowed by quantum mechanics is different from the conventional concept of correlation. The paradoxical concept of quantum correlation started from Einstein, Podolsky and Rosen (EPR) who questioned the completeness of quantum mechanics based on local realism in their seminal paper~\cite{epr}.  It seems that in this question, Einstein was not fully content with the indeterministic nature of quantum theory.  A class of theories called local hidden-variable theories emerged in the search of a complete theory.  About 30 years after the EPR paper, Bell~\cite{bell} came up with an experimentally testable inequality which a local hidden-variable theory should satisfy.   Clauser, Horne, Shimony and Holt (CHSH)~\cite{chsh}'s version of Bell's inequality for a spin-${1\over 2}$ system is
\begin{equation}
\langle \hat{A}\hat{B}\rangle+ \langle \hat{A}^\prime\hat{B}\rangle+ \langle \hat{A}\hat{B}^\prime\rangle- \langle \hat{A}^\prime\hat{B}^\prime\rangle\leq 2
\label{chsh}
\end{equation}
where $\hat{A},\hat{A}^\prime$ are two observables, taking values either 1 or $-1$, for one system and $\hat{B}, \hat{B}^\prime$ are two other observables for the other system.
However, for the singlet state
\begin{equation}
|\psi_{sing}\rangle={1\over \sqrt{2}}(|\uparrow\rangle_a |\downarrow\rangle_b- |\downarrow\rangle_a |\uparrow\rangle_b),
\label{singlet}
\end{equation}
the correlation function on the left hand side of the inequality can be as large as $2\sqrt{2}$, which violates the local hidden variable theory.  This was experimentally proven using the two polarisation states of photons, instead of ups and downs of spins (See a recent experimental result~\cite{vienna} for some hints and problems in the experimental proof of the violation).

Differently from the spin system, which can have two measurement outcomes of spin up and spin down, the light field we consider in the paper is defined in an infinite dimensional Hilbert space.  For instance, the number of photons in the field can be $0,1,2, \cdots, \infty$.  In phase space, the light field is described by a continuous variable.  There have been efforts to consider the nonlocality of light fields, using dichotomic~\cite{kim-jeong} and multichotomic~\cite{kim-son} observables as well as continuous-variable correlations~\cite{reid}.

Apart from the test of nonlocality based on the EPR paradox and Bell's inequality, the concept of entanglement has been extensively studied as  one of the key resources for the current development of quantum information processing.  While the test of nonlocality is using physical observables, entanglement started as a mathematical concept.  When a composite system of two subsystems a and b is not entangled (separable), the density $\hat{\rho}$ of the composite system can be written as a statistical sum
\begin{equation}
\hat{\rho}=\sum_{i}p(i)\hat{\rho}_i^a\hat{\rho}_i^b
\label{separable}
\end{equation}
where $p(i)$ is a probability function and $\hat{\rho}_i^a$ and $\hat{\rho}_i^b$ are density operators for subsystems a and b, respectively.  A density operator is a positive normalised operator\footnote{A positive operator is defined as an operator whose eigenvalues are all positive.} and its transposition is again a density operator.  Thus a partial transposition,
\begin{equation}
\hat{\rho}^{\mbox{\tiny{PT}}}=\sum_{i}p(i)[\hat{\rho}_i^a]^{\mbox{\tiny{T}}}\hat{\rho}_i^b,
\label{PPT}
\end{equation}
of a separable density operator (\ref{separable}) has to be
positive, where the superscripts {\tiny PT} and {\tiny T} denote
partial transposition and transposition\footnote{For the purpose of
an entanglement test or an entanglement measure, the partial
transposition of either of the two modes will bear the same
result.}.  Thus, using the converse, we can say that the negativity
of the partially transposed density operator is a sufficient
condition for the entanglement of the composite system.  The
Horodecki family proved that the negative partial transposition (NPT) condition is also a necessary condition for
the entanglement of certain classes of composite systems~\cite{horodecki}.  A few
years later, Lee {\it et al.} proved that the NPT is an
entanglement monotone thus can be used as a measure of entanglement
for a spin-${1\over 2}$ system~\cite{lee-kim}, which was extended to
a high-dimensional system by Vidal and Werner~\cite{vw}.  For a
Gaussian two-mode field, the NPT is also a sufficient and
necessary condition~\cite{simon} for entanglement.  On the other
hand, for a non-Gaussian two-mode field, a more elaborative
entanglement condition applies~\cite{vogel-e}.

Quantum entanglement is a mathematically well-defined concept and quite universal in the sense that entanglement in a composite system means its possession of quantum-mechanical correlation, regardless of its usefulness for a particular task.  Quantum teleportation, on the other hand, is a physical task for which entanglement of a quantum channel is necessary for its success.  Even though quantum entanglement is not a sufficient condition, the success of quantum teleportation can be regarded as a physically measurable witness of entanglement.  In this sense, the nonlocality test discussed earlier is another experimentally accessible entanglement witness.

After the two particles of a perfect entangled pair are shared between two remote stations, the quantum state to be teleported and one particle of the entangled pair are jointly measured at one station.  The other particle of the entangled pair at the other station is then collapsed into  a state.   The original quantum information can be recovered by a unitary transformation according to the classical message on the measurement outcome (See Ref.~\cite{braunstein01} for a review on quantum teleportation of a coherent state and continuous variables.).

\subsection{Teleportation with a Photon-subtracted squeezed state}
For a two-mode (modes $a$ and $b$) continuous-variable system defined in an infinite-dimensional Hilbert space, a maximally entangled state is
\begin{equation}
|\Psi_{max-un}\rangle=\sum_{n=0}^\infty|n\rangle_a|n\rangle_b
\label{max-un}
\end{equation}
which is unnormalised and the energy is infinite hence the state is unphysical.  On the other hand, the two-mode squeezed state $|{\cal S}_2\rangle$ in Eq.~(\ref{sq2}) is entangled with a finite degree.  The degree of entanglement for a pure state can be obtained by the von Neumann entropy of the marginal density operator of a subsystem.  For the two-mode squeezed state of density operator $\hat{\rho}_{|{\cal S}_2\rangle}$, the degree of entanglement $\mathfrak{E}$ is
\begin{equation}
\mathfrak{E}(\hat{\rho}_{|{\cal S}_2\rangle})=-\mbox{Tr}_a[\hat{\rho}_a\log_2\hat{\rho}_a]
\label{von-N}
\end{equation}
where
\begin{equation}
\hat{\rho}_a=\mbox{Tr}_b\left(\hat{\rho}_{|{\cal S}_2\rangle}\right)
={1\over\cosh^2\zeta}\sum_{n=0}^\infty\tanh^{2n}\zeta|n\rangle\langle
n|.
\label{marginal}
\end{equation}
The marginal density operator  of the two-mode squeezed
state is already diagonalised so that the calculation is rather
straightforward but for a general two-mode state, diagonalisation
may be a challenging task.  On the other hand,
$\mathfrak{M}(\hat{\rho})= 1-\mbox{Tr}_a\hat{\rho}_a^2$, based on a
linearised entropy, can be more easily calculated even though it
does not provide statistical information as well as the von Neumann
entropy does.

Because of the experimental limitations, a highly entangled state is
difficult to obtain.  For example, in the first demonstration of the
continuous-variable quantum teleportation~\cite{furusawa}, the
squeezing was low hence the low entanglement of the quantum channel and
the average fidelity of quantum teleportation was $58\pm2\%$, which
is just above the classical limit of 50\%. In order to improve the
fidelity of quantum teleportation, Opatrn\'{y}, Kurizki and
Welsch~\cite{opatrny} suggested to select an optimal subensemble of
entangled fields based on a conditional measurement, which is in
effect subtracting a photon from each mode of the two-mode squeezed
state.  Once the quantum channel is prepared in this way, the rest
of the quantum teleportation protocol is performed.  For the initial
state of a coherent superposition state, they show the improvement
of the teleportation fidelity. At the end of their investigation,
Opatrn\'{y} and colleagues briefly discuss that they do not find an
improvement of quantum teleportation for the quantum channel
prepared by adding a definite number of photons to each mode of the
two-mode squeezed state. Olivares, Paris and Bonifacio~\cite{oliv}, on the other hand, considered the photon subtraction using on-off photodetectors and showed the improvement of quantum teleportation depending on various parameters involved.

\begin{figure}
\centering{\includegraphics{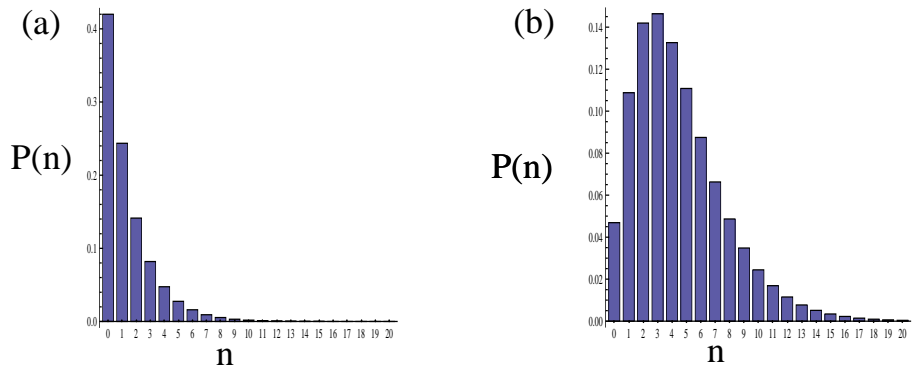}}
\caption{Photon
number distributions  for mode $a$ of a two-mode squeezed state before {\bf
(a)} and after {\bf (b)} subtracting a photon from each mode.
Squeezing parameter $\zeta=1$.} \label{fig:photon-number}
\end{figure}

The von Neumann entropy  measure $\mathfrak{E}$ of entanglement for
the two-mode squeezed state when the squeezing parameter $\zeta=1$
is 2.34 while that is increased to 3.53 for the photon-subtracted
squeezed state.  The linearised entropy measure
$\mathfrak{M}(\hat{\rho})$ of entanglement also shows that the
photon-subtracted squeezed state is more highly entangled than the
original squeezed state for non-zero squeezing:
\begin{equation}
\mbox{Tr}_a\left(\mbox{Tr}_{b}\hat{\rho}_{|{\cal S}_2\rangle}\right)^2 > \mbox{Tr}_a(\hat{\rho}_{a}^{ts})^2
\label{entropy-com}
\end{equation}
where  $\hat{\rho}_a^{ts}=\mbox{Tr}_{b}\hat{\rho}^{ts}$ and $\hat{\rho}^{ts}$ is the density operator for
the field after a photon is subtracted from each mode of the
two-mode squeezed vacuum,
\begin{equation}
\hat{\rho}_{a}^{ts}=\frac{1}{\cosh^4\zeta\cosh2\zeta}
\sum_{n=0}^\infty
(\tanh\zeta)^{2n} (n+1)^2|n\rangle_a\langle n|. \label{2sq-sub}
\end{equation}
Cochrane, Ralph and Milburn~\cite{cochrane} calculated the photon number distribution $P(n)$ for the marginal state $\hat{\rho}_a^{ts}$:
\begin{equation}
P(n)=\langle
n|\hat{\rho}_a^{ta}|n\rangle=
\frac{1}{\cosh^4\zeta\cosh2\zeta}\tanh^{2n}\zeta
(n+1)^2 \label{p-n-2sq}
\end{equation}
As can be seen clearly in Fig.~\ref{fig:photon-number}, the photon
number distribution is peaked sharply at $n=0$ for the two-mode
squeezed state while for the single-photon subtracted squeezed
state, the peak is moved to $n=3$ and the width of the distribution
is wider with a smooth peak.

Entanglement has two important ingredients: random and deterministic
nature. As can be seen in the singlet state (\ref{singlet}), before
the measurement a subsystem is equally likely be measured in spin up or down; hence the random nature. However, once the
measurement is performed on a subsystem, the state of the other
subsystem collapses into an eigenstate of the measurement so that we
can predict the measurement outcome without measurement; hence the
deterministic nature. In order to further the analysis, let us
consider a pure entangled state
$|\psi_{gen}\rangle=c_0|\uparrow\rangle_a |\downarrow\rangle_b-
c_1|\downarrow\rangle_a|\uparrow\rangle_b$ with real $c_0$ and
$c_1$. When $c_0=c_1$, the outcome of the first measurement is
maximally random. Now, we perform local unitary transformations
$\hat{U}$ to rotate spins in such a way,
$$
|\uparrow\rangle\stackrel{\hat{U}}{\longrightarrow}{1\over\sqrt{2}}(|\uparrow\rangle+|\downarrow\rangle)~~,~~
|\downarrow\rangle\stackrel{\hat{U}}{\longrightarrow}{1\over\sqrt{2}}(|\uparrow\rangle-|\downarrow\rangle).
$$
The $|\psi_{gen}\rangle$ turns into
$$
|\psi_{gen}^\prime\rangle={1\over2}|\uparrow\rangle_a[(c_0-c_1)|\uparrow\rangle- (c_0+c_1)|\downarrow\rangle]_b + {1\over2}|\downarrow\rangle_a[(c_0+c_1)|\uparrow\rangle- (c_0-c_1)|\downarrow\rangle]_b.
$$
Two states $(c_0\mp c_1)|\uparrow\rangle- (c_0\pm
c_1)|\downarrow\rangle$ for  particle b are not orthogonal to each
other unless $c_0^2=c_1^2$.  Because of the ambiguity due to
non-orthogonality, the correlation between particles a and b is not
as clear as for the singlet case ($|\psi_{gen}\rangle$ when
$c_0=c_1$).  Extending this idea, we can say that the enhancement of
entanglement is related to the flatter weights of each component
states in the photon subtracted squeezed state\footnote{This should
not be generalised too much but at least this gives an intuitive
view into the problem.}.

As discussed earlier in Eq.~(\ref{on-off}), the experimental
realisation of single-photon measurements is approximated by an
on-off photodetector. In this case, once a measurement is done on
one mode, the field of the other mode is collapsed into a mixed
state. Kitagawa {\it et al.} studied the degree of entanglement for
the photon subtracted two-mode squeezed state with this experimental
consideration~\cite{kitagawa}.

\subsection{Nonlocality of a photon-subtracted squeezed state}
While the original EPR paradox was on a continuous-variable
system, Bell, CHSH and many followers developed inequalities for the
so-called Bohm's version of the EPR state, which is for a
spin-${1\over 2}$ system.  This system has only two measurement
outcomes so that it is much easier to treat than the
infinite-dimensional continuous variables. (There have been
discussions on the validity of a nonlocality test for a
multichotomic, many observables and many-dimensional system.  As
this is beyond the scope of the present paper, we will refer to
Ref.~\cite{kim-son} and references there in for further studies.)

Bell wrote that the EPR state,  which is simulated by a two-mode
squeezed state when the squeezing parameter is infinity, would not
violate his inequality because its Wigner function is positive
everywhere in phase space.  Using the fact that a Gaussian state
transforms into a non-Gaussian state by subtracting a photon, Nha
and Carmichael~\cite{nha} suggested the CHSH Bell's inequality test
for a two-mode squeezed state, conditioned on subtraction of a
photon from each mode of the state.  Their dichotomic observable is
based on a homodyne measurement, which has high detection efficiency
to avoid a detection loophole in an experimental nonlocality test.
As shown in Eq.~(\ref{homodyne-theta}), the phase value is measured
by a homodyne detector.  The observable setting is determined by the
phase of a local oscillator ($\phi_L$ in
Eq.~(\ref{homodyne-theta})).

For the dichotomic value, +1 is endorsed when the measurement
outcome is positive and -1 otherwise. In their analysis, Nha and
Carmichael considered the realistic way to subtract a photon using a
beam splitter and an on-off photodetector. They found that the CHSH
Bell's inequality is violated even when the squeezing parameter
$\zeta$ is as small as 0.48 as far as the transmittivity of the beam
splitter is high ($\approx 1$).  They also found that the scheme is
quite insensitive to the detection efficiency of the on-off
detector.  It is because if a photon is not detected, they will
anyway throw away the data so losing a photon is not a problem.  In
the same sense, losing a photon does not cause a problem in the
on-off photodetector. Nha and Carmichael
carefully remark against the possible fallacy due to the selection
of data as the test is conditioned on subtraction of a photon.  They
say~\cite{nha} `{\it the situation differs from the so-called
detection loophole which asserts that unfair data sampling can lead
to the violation of a Bell inequality even for a classcially
correlated state.}'

Garc\'{i}a-Patr\'{o}n {\it et al.}~\cite{garcia} also studied the
same system and assert that the local hidden variable theory does
not change at all by the subtraction process because it cannot be
influenced by the two local homodyne measurements. They considered
the detection efficiency of the homodyne detectors and found 1\% of
CHSH Bell's inequality violation for the homodyne efficiency of
about 95\%.  In a longer version of this work,
Garc\'{i}a-Patr\'{o}n, Fiur\'{a}sek and Cerf~\cite{garcia} extended
the scheme to consider various scenarios to subtract photons using
beam splitters and on-off photodetectors. They found that the
violation would increase by subtracting one, two and four photons
from each mode but no violation occurs by subtracting three photons.

\subsection{Experiment: entanglement enhancement by single counting}
Despite theoretical studies on increasing nonlocality  violation
and teleportation fidelity, experimental realisation has not been
made to subtract photons from each mode of the field.  It is because
the probability of subtracting a photon from one mode is already
very rare and difficult so that coincidence subtraction of two
photons is  technically extremely demanding. However, there is a
scheme which is less demanding because only one photon subtraction
is involved to increase entanglement.  If the subtraction is ideal
for the operation to be represented simply by an annihilation
operator, a single-photon subtraction from one mode of the two-mode
squeezed state will increase its entanglement. The two-mode squeezed
state turns into
\begin{equation}
|{\cal S}_{2s}\rangle={2\over \sinh2\zeta}\sum_{n=1}^\infty \sqrt{n}(-\tanh\zeta)^n |n-1\rangle_a|n\rangle_b.
\end{equation}
The linearised entropy entanglement measure for this is
\begin{equation}
\mathfrak{M}(\hat{\rho}_{2s})=1-\left(\frac{2}{\sinh
2\zeta}\right)^4\sum_{n=1}^\infty n^2\tanh^{4n}\zeta
\label{linear-2s}
\end{equation}
which is higher than its counter part for the initial two-mode squeezed state for all nonzero squeezing parameter $\zeta$.

Ourjoumtsev {\it et al.}~\cite{our} use a technique to subtract only
one photon but by erasing the information on the path information of
the subtracted photon, they increase the degree of
entanglement. Let us consider a scheme to annihilate a photon from
one of the modes of a two-mode field.  Then the operation is
described by $(\hat{a}+\hat{b})/\sqrt{2}$.  After this
operation, the two-mode squeezed state $|{\cal S}_2\rangle$ turns
into
\begin{equation}
|{\cal S}_{2e}\rangle=\frac{\sqrt{2}}{\mbox{sinh}2\zeta}\sum_{n=1}^\infty \sqrt{n}(-\tanh{\zeta})^n(|n\rangle_a|n-1\rangle_b +|n-1\rangle_a |n\rangle_b)
\label{s-2e}
\end{equation}
The marginal density operator for mode $a$ can be  calculated as
$\hat{\rho}_a=\mbox{Tr}_b|{\cal S}_{2e}\rangle\langle {\cal
S}_{2e}|$:
\begin{eqnarray}
&&\hat{\rho}_a=\frac{2}{\sinh^22\zeta}\sum_{n=1}^\infty n\tanh^{2n}\zeta\left[|n\rangle\langle n| +\frac{1}{\tanh\zeta}\sqrt{\frac{n+1}{n}}|n+1\rangle\langle n-1| \right.
\nonumber \\
&&\left. +\tanh\zeta\sqrt{\frac{n+1}{n}}|n-1\rangle\langle n+1| +|n-1\rangle\langle n-1| \right]_a
\end{eqnarray}
whose $\mathfrak{M}$ is larger than $\mathfrak{M}(\hat{\rho}_{|{\cal S}_2\rangle})$.
\begin{figure}
\centering{\includegraphics{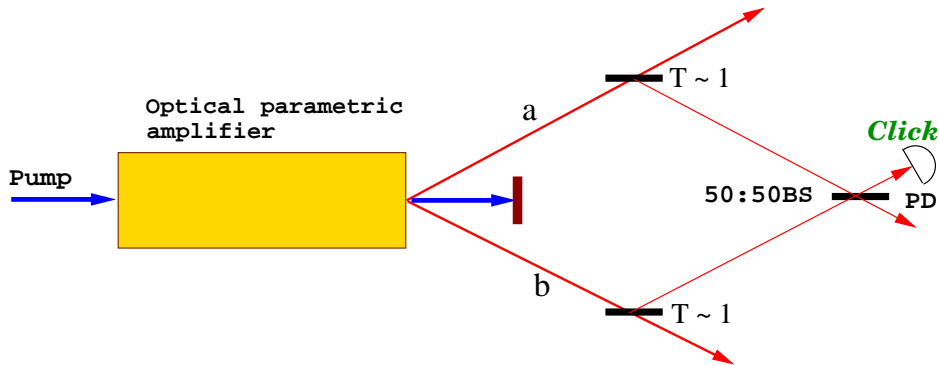}}
\caption{Ourjoumtsev {\it et al.}}~\cite{our} increases the degree of entanglement by subtracting one photon from one of the mode of a two-mode squeezed state.  PD: photodetector, and T: transmittance of the beam splitter.
\label{fig:our}
\end{figure}

With the experimental setup as in Fig.~\ref{fig:our},  Ourjoumtsev
{\it et al.} show an enhancement of entanglement for small
squeezing.   In the experiment, as an on-off photodetector is used,
the resultant state in the experiment is mixed.  Thus the von
Neumann entropy or the linear entropy cannot be used to measure the
entanglement.  They reconstruct the density matrix from the homodyne
measurements, then using the entanglement monotone based on the NPT
condition~\cite{vw} they show that the photon-subtracted state has
higher entanglement

\section{Final remarks}
Since  its advent, a laser light field has been an important tool to
provide experimental evidences of paradoxical ideas in quantum
physics.  In the current development of quantum information
processing and quantum control of a system, it is very useful to
generate a light field in an arbitrary quantum state.  In this
paper, we have revised how to engineer the quantum state of a
travelling light field using beam splitters, parametric amplifiers,
photodetectors and homodyne detectors.  We have seen that
single-photon subtraction and addition are possible and these have
been demonstrated experimentally.  It is also possible to increase
entanglement of a two-mode light field using the same technique.

Even though a sequence of subtraction or addition enables to engineer a quantum state to an arbitrary state, it is experimentally demanding. It will be interesting to have theoretical suggestions which may be easily realisable by experiments for tests of quantum physics and to increase the applicability of quantum optics.

\subsection*{Acknowledgments} I acknowledge financial support from the UK EPSRC and QIP IRC.
I would like to thank M. Bellini,  J. Fiur\'{a}\v{s}ek, G. Gribakin, H. Jeong, H. W. Lee,  J. Lee, J. F. McCann, P. Marek, W. Son,  E. Park,  M. Sasaki, A. Zavatta for discussions and comments.



\section*{References}
\begin{thebibliography}{99}
\bibitem{bose} S. N. Bose, Z. Phys. {\bf 26}, 178 (1924).
\bibitem{einstein} A. Einstein, Sitz. Ber. Preuss. Akad. Wiss. (Berlin) {\bf 1}, 3
(1925).
\bibitem{dirac} P. A. M. Dirac, The Principles of Quantum Mechanics (Oxford University Press, Oxford, 1957).
\bibitem{ballentine} L. E. Ballentine, Quantum Mechanics: A Modern Develpment (World Scientific, Singapore, 1998).
\bibitem{glauber01} R. J. Glauber, Phys. Rev. {\bf 131}, 2766 (1063).
\bibitem{glauber02} K. E. Cahill and R. J. Glauber, Phys. Rev. {\bf 177}, 1857 (1969); {\it ibid}, 1882 (1969).
\bibitem{wigner} E. P. Wigner, Phys. Rev. {\bf 40} 749 (1932).
\bibitem{glauber03} R. J. Glauber, Phys. Rev. Lett. {\bf 10}, 84 (1963).
\bibitem{molmer} K. M{\o}lmer, Phys. Rev. A {\bf 55}, 3195 (1997); T. Rudolph and B. C. Sanders, Phys. Rev. Lett. {\bf 87}, 077903 (2001).
\bibitem{bu-knight} V. Bu\v{z}ek and P. L. Knight, Progress in Optics XXXIV, edited by E. Wolf, p. 1 (Elsevier, Amsterdam, 1995).
\bibitem{caves1}C. M. Caves and B. L. Schumaker, Phys. Rev. A {\bf 31}, 3068 (1985); {\it ibid} 3093 (1985).
\bibitem{barnett-radmore} S. M. Barnett and P. M. Radmore, Methods in Theoretical Quantum Optics (Oxford University Press, Oxford, 1997)
\bibitem{kim01} M. S. Kim, W. Son, V. Bu\v zek and P. L. Knight, Phys. Rev. A {\bf 65} 032323 (2002).
\bibitem{braunstein01} S. L. Braunstein and P. van Loock, Rev. Mod. Phys. {\bf 77}, 513 (2005).
\bibitem{huang} H. Huang and G. S. Agarwal, Phys. Rev. A {\bf 49}, 52 (1994).
\bibitem{campos} R. A. Campos, B. E. A. Saleh and M. C. Teich, Phys. Rev. A {\bf 40}, 1371 (1989).
\bibitem{yuen}H. P. Yuen and J. H. Shapiro, IEEE Trans. Inf. Theory {\bf 24}, 657 (1978); J. H. Shapiro, H. P. Yuen, M. J. A. Machado, IEEE Trans. Inf. Thoery {\bf 25}, 179  (1979).
\bibitem{hong} C. K. Hong, Z. Y. Ou and L. Mandel, Phys. Rev. Lett. {\bf 59}, 2044 (1987).
\bibitem{vogel} K. Vogel and H. Risken, Phys. Rev. A {\bf 40}, 2847 (1989).
\bibitem{yuen-ineff} H. P. Yuen and V. W. S. Chen, Opt. Lett. {\bf 8}, 177 (1983).
\bibitem{braunstein-hom} S. L. Brauntstein, Phys. Rev. A {\bf 42}, 474 (1990).
\bibitem{ulf2} U. Leonhardt, Measuring the Quantum State of Light (Cambridge University Press, Cambridge, 1997).
\bibitem{sch} E. Schr\"{o}dinger, Naturwissenschaften {\bf 23}, 807, 823, 844 (1935).
\bibitem{agarwal-quasi} G. S. Agarwal and E. Wolf, Phys. Rev. D {\bf 2}, 2161 (1970); {\it ibid}, 2182 (1970).
\bibitem{sudarshan} E. C. Sudarshan, Phys. Rev. Lett. {\bf 10}, 277 (1963).
\bibitem{loudon} R. Loudon, The Quantum Theory of Light, 2nd edition (Clarendon Press, Oxford, 1983).
\bibitem{barnett-n} N. L\"{u}tkenhaus and S. M. Barnett, Phys. Rev. A {\bf 51}, 3340 (1995).
\bibitem{kim-imoto} M. S. Kim and N. Imoto, Phys. Rev. A {\bf 52}, 2401 (1995).
\bibitem{grangier01} J. Wenger, R. Tualle-Brouri, and P. Grangier, Phys. Rev. Lett. {\bf 92}, 153601 (2004).
\bibitem{bellini01} A. Zavatta, S. Viciani and M. Bellini, Science {\bf 306}, 660 (2004).
\bibitem{bellini-kim} V. Parigi, A. Zavatta, M. S. Kim and M. Bellini, Science {\bf 317}, 1890 (2007); R. W. Boyd, K. W. Chan, M. N. O'Sullivan, Science {\bf 317}, 1874(2007).
\bibitem{knight-jcm} B. W. Shore and P. L. Knight, J. Mod. Opt. {\bf 40}, 1195 (1993).
\bibitem{haroche-single} P. Bertet, A. Auffeves, P. Maioli, S. Osnaghi, T. Meunier, M. Brune, J. M. Raimond, and S. Haroche, Phys. Rev. Lett. {\bf 89}, 200402 (2002).
\bibitem{polzik01} J. S. Neergaard-Nielsen, B. Melholt Nielsen, C. Hettich, K. M{\o}lmer
and E. S. Polzik, Phys. Rev. Lett. {\bf 97}, 083604 (2006); K. Wakui, H. Takahashi, A. Furusawa and M. Sasaki, Opt. Exp. {\bf 15}, 3568 (2007).
\bibitem{grangier02} A. Ourjoumtsev, R. Tualle-Brouri, J. Laurat and P. Grangier, Science
{\bf 312}, 83  (2006).
\bibitem{schwinger} J. Schwinger, in `Quantum Theory of Angular Momentum' edited by L. C. Biedenharn and H. van Dam (Academic, New York, 1965).
\bibitem{kim-mc} H. McAneney, J. Lee and M. S. Kim, Phys. Rev. A {\bf 68}, 063814 (2003).
\bibitem{pegg-barnett} D. T. Pegg, L. S. Phillips and S. M. Barnett, Phys. Rev. Lett. {\bf 81}, 1604 (1998).
\bibitem{wheeler} translation of \cite{sch} in Quantum Theory and Measurement, edited by J. A. Wheeler and W. H. Zurek (Princeton university Press, New Jersey 1983).
\bibitem{jeong-kim} H. Jeong, M. S. Kim, T. C. Ralph and B. S. Ham, Phys. Rev. A {\bf 70}, 061801 (2004).
\bibitem{dakna99} M. Dakna, J. Clausen, L. Kn\"o ll and D.-G. Welsch, Phys. Rev. A {\bf 59}, 1658 (1999).
\bibitem{paris} M. G. A. Paris, Phys. Lett. A {\bf 217}, 78 (1996).
\bibitem{fiu1} J. Fiur\'{a}\v{s}ek, R. Garc\'{i}a-Patr\'{o}n and N. J. Cerf, Phys. Rev. A {\bf 72}, 033822 (2005).
\bibitem{dakna97} M. Dakna, T. Anhut, T. Opatrn\'{y}, L. Kn\"{o}ll and D.-G. Welsch, Phys. Rev. A {\bf 55}, 3184 (1997).
\bibitem{kim-munro} M. S. Kim, J. Lee and W. J. Munro, Phys. Rev. A {\bf 66}, 030301(R) (2002).
\bibitem{kim-lee} J. Lee, M. S. Kim and \v{C}. Brukner, Phys. Rev. Lett. {\bf 91}, 0807902 (2003).
\bibitem{kim-jeong-park} M. S. Kim, E. Park, P. L. Knight and H. Jeong, Phys. Rev. A {\bf 71}, 043805 (2005).
\bibitem{molmer02}K. M{\o}lmer, Phys. Rev. A {\bf 73}, 063804 (2006).
\bibitem{knight-barry} B. M. Garraway and P. L. Knight, Phys. Rev. A {\bf 50}, 2548 (1994).
\bibitem{grangier03} A. Ourjourmtsev, H. Jeong, R. Tualle-Brouri and P. Grangier, Nature {\bf 448}, 784 (2007).
\bibitem{grangier04} A. Ourjoumtsev, R. Tualle-Brouri and P. Grangier, Phys. Rev. Lett. {\bf 96}, 213601 (2006).
\bibitem{agarwal-tara} G. S. Agarwal and K. Tara, Phys. Rev. A {\bf 43}, 492 (1991).
\bibitem{lee} C. T. Lee, Phys. Rev. A {\bf 52}, 3374 (1995).
\bibitem{bellini02} A. Zavatta, V. Parigi and M. Bellini, Phys. Rev. A {\bf 75}, 052106 (2007).
\bibitem{mandel} L. Mandel, Opt. Lett. {\bf 4}, 205 (1979).
\bibitem{epr} A. Einstein, B. Podolsky, and N. Rosen, Phys. Rev. {\bf 47} 777 (1935).
\bibitem{bell} J.S. Bell, Physics {\bf 1}, 195 (1964).
\bibitem{chsh} J. F. Clauser, M. A. Horne, A. Shimony and R. A. Holt, Phys. Rev. Lett. {\bf 23}, 880 (1969).
\bibitem{vienna} S. Gr\"oblacher, T. Paterek, R. Kaltenbaek, \v{C} . Brukner, M. \.{Z}ukowski,
M. Aspelmeyer and A. Zeilinger, Nature {\bf 446}, 871 (2007).
\bibitem{kim-jeong} H. Jeong, W. Son, M. S. Kim, D. Ahn and \v{C}. Brukner, Phys. Rev. A {\bf 67}, 012016 (2003)
\bibitem{kim-son} W. Son, \v{C}. Brukner and M. S. Kim, Phys. Rev. Lett. {\bf 97}, 110401 (2006); W. Son, J. Lee and M. S. Kim, Phys. Rev. Lett. {\bf 96}, 060406 (2006).
\bibitem{reid} E. G. Cavalcanti, C. J. Foster, M. D. Reid and P. D. Drummond, Phys. Rev. Lett. {\bf 210405} (2007).
\bibitem{horodecki} M. Horodecki, P. Horodecki and R. Horodecki, Phys. Lett. A {\bf 223}, 1 (1996).
\bibitem{lee-kim} J. Lee, M. S. Kim, Y. J. Park and S. Lee, J. Mod. Opt. {\bf 47}, 2151 (2000); J. Lee and M. S. Kim, Phys. Rev. Lett. {\bf 84}, 4236 (2000).
\bibitem{vw} G. Vidal and R. F. Werner, Phys. Rev. A {\bf 65}, 032314 (2002).
\bibitem{simon} R. Simon, Phys. Rev. Lett. {\bf 84}, 2726 (2000).
\bibitem{vogel-e} G. S. Agarwal and A. Biswas, New J. Phys. {\bf 7}, 211 (2005); E. Schshkin and W. Vogel, Phys. Rev. Lett. {\bf 95}; 230502 (2005); M. Hillery and M. S. Zubairy, Phys. Rev. Lett. {\bf 96}, 050503 (2006);
\bibitem{furusawa} A. Furusawa, J. L. S{\o}rensen, S. L. Braunstein, C. A. Fuchs, H. J. Kimble and E. S. Polzik, Science {\bf 282}, 706 (1998).
\bibitem{opatrny} T. Opatrn\'{y}, G. Kurizki and D.-G. Welsch, Phys. Rev. A {\bf 61}, 032302 (2000).
\bibitem{oliv} S. Olivares, M. G. A. Paris and R. Bonifacio, Phys. Rev. A {\bf 67}, 032314 (2003).
\bibitem{cochrane} P. T. Cochrane, T. C. Ralph and G. J. Milburn, Phys. Rev. A {\bf 65}, 062306 (2002).
\bibitem{kitagawa} A. Kitagawa, M. Takeoka, M. Sasaki and A. Chefles, Phys. Rev. A {\bf 73}, 042310 (2006).
\bibitem{nha} H. Nha and H. J. Carmichael, Phys. Rev. Lett. {\bf 93}, 020401 (2004).
\bibitem{garcia} R. Garc\'{i}a-Patr\'{o}n, J. Fiur\'{a}\v{s}ek, N. J. Cerf, J. Wenger, R. Tualle-Brouri and Ph. Granger, Phys. Rev. Lett. {\bf 93}, 130409 (2004); R. Garc\'{i}a-Patr\'{o}n, J. Fiur\'{a}\v{s}ek and N. J. Cerf, Phys. Rev. A {\bf 71}, 022105 (2005).
\bibitem{our} A. Ourjoumtsev, A. Dantan, R. Tualle-Brouri and P. Grangier, Phys. Rev. Lett. {\bf 98}, 030502 (2007).
\end{thebibliography}
\end{document}